\documentclass[12pt, letter]{article}

\renewcommand{\baselinestretch}{1.75}

\usepackage{xr}
\usepackage{amsthm,amsmath,amssymb,bbm}
\usepackage{natbib}
\usepackage{multirow}
\usepackage[pdftex]{graphicx}
\usepackage{subfigure}
\usepackage{booktabs}
\usepackage{algorithm}
\usepackage{algorithmic}
\usepackage{lipsum}
\usepackage{mathrsfs}
\usepackage{dsfont}
\usepackage{titling}
\usepackage{amsfonts}       
\usepackage{nicefrac}       
\usepackage{microtype}      
\usepackage{color}
\usepackage{comment}
\usepackage{enumerate}
\usepackage{setspace}
\usepackage[usenames,dvipsnames,svgnames,table]{xcolor}
\usepackage[colorlinks,
linkcolor=red,
anchorcolor=blue,
citecolor=blue
]{hyperref}

\usepackage{smile}

\renewcommand{\hat}{\widehat}
\newcommand{\bfm}[1]{\ensuremath{\mathbf{#1}}}
\newcommand{\prob}{\text{Pr}}
\newcommand{\red}{\textcolor{red}}

\def\bff{\bfm f}    
   \def\bG{\bfm G}  
\def\bh{\bfm h}     
   \def\bI{\bfm I}

\def\bm{\bfm m}

     \def\RR{\mathbb{R}}

\def\bu{\bfm u}     
\def\bv{\bfm v}     
\def\bw{\bfm w}     
\def\bx{\bfm x}   \def\bX{\bfm X}  
\def\by{\bfm y}   \def\bY{\bfm Y}

\def\calA{{\cal  A}} 
\def\calB{{\cal  B}}

\def\calE{{\cal  E}} 
\def\calF{{\cal  F}}

\def\calS{{\cal  S}}

\def\calW{{\cal  W}}

\def\E{\text{E}}
\def\R{\mathbb{R}}
\def\cov{\mathrm{cov}}
\def\pcov{\mathrm{Pcov}}
\def\pc{\mathrm{PC}}
\def\acos{\mathrm{arccos}}
\def\FDR{\mathrm{FDR}}

\newcommand{\bfsym}[1]{\ensuremath{\boldsymbol{#1}}}

\def\balpha{\bfsym \alpha}
\def\bbeta{\bfsym \beta}

\def\bmu{\bfsym {\mu}}                 

\def\btheta{\bfsym {\theta}}           
          
             \def\bSigma{\bfsym \Sigma}

 \def\homega {\hat {\omega}}

\def \T{\mathrm{\scriptstyle T}} 
\def\FDP{\mathrm{FDP}}
\def\wt{\widetilde}
\newcommand\independent{\protect\mathpalette{\protect\independenT}{\perp}}
\def\independenT#1#2{\mathrel{\rlap{$#1#2$}\mkern2mu{#1#2}}}
\newcommand{\Rom}[1]{\text{\uppercase\expandafter{\romannumeral #1\relax}}}
\def\singlespace{\def\baselinestretch{1}}

\renewcommand{\baselinestretch}{1.1}

\AtBeginDocument{}

\makeatletter
\def\singlespace{\def\baselinestretch{1}\@normalsize}

\renewcommand{\baselinestretch}{1.0}

\begin{document}



\title{Model-free Feature Screening and \\ FDR Control with Knockoff Features}

\author{Wanjun Liu\thanks{Department of Statistics, The
Pennsylvania State University, University Park, PA 16802. E-mail:
\href{mailto:wxl204@psu.edu}{\textsf{wxl204@psu.edu}}.},~~Yuan
Ke\thanks{Department of Statistics,  University of Georgia, Athens, GA 30602.
E-mail: \href{mailto:yuan.ke@uga.edu}{\textsf{yuan.ke@uga.edu}}.}, Jingyuan Liu
\thanks{MOE Key Laboratory of Econometrics, Department of Statistics,
School of Economics, Wang Yanan Institute for Studies in Economics,
and Fujian Key Lab of Statistics, Xiamen University, P.R China.
E-mail:
\href{mailto:jingyuan@xmu.edu.cn}{\textsf{jingyuan@xmu.edu.cn}}}
~~and~~Runze Li\thanks{Department of Statistics, The Pennsylvania State
University, University Park, PA 16802. E-mail:
\href{mailto:rzli@psu.edu}{\textsf{rzli@psu.edu}}.} }

\date{}
\maketitle
\vspace{-0.5in}

\begin{abstract}
This paper proposes a model-free and data-adaptive feature screening method for ultra-high dimensional data. The proposed method is based on the projection correlation which measures the dependence between two random vectors. This projection correlation based method does not require specifying a regression model, and applies to data in the presence of heavy tails and multivariate responses. It enjoys both sure screening and rank consistency properties under weak assumptions.  A two-step approach, with the help of knockoff features, is advocated to specify the threshold for feature screening  such that
the false discovery rate (FDR) is controlled under a pre-specified level. The proposed two-step approach enjoys both sure screening and FDR control simultaneously if the pre-specified FDR level is greater or equal to $1/s$, where $s$ is the number of active features.  The superior empirical performance of the proposed method is illustrated by simulation examples and real data applications.
\end{abstract}


\baselineskip=25pt

\renewcommand{\baselinestretch}{1.75}
\section{Introduction}
\label{sec:intro}

The technological development has made extensive data collection and storage feasible in diverse fields. Datasets with ultra-high dimensional features characterize many contemporary research problems in machine learning, computer science,  statistics,  engineering, social science, finance and so on.
When the features contain redundant or noisy information, estimating their functional relationship with the response may become quite challenging in terms of computational expediency, statistical accuracy, and algorithmic stability \citep{fan2009ultrahigh, hall2009using, lv2014model}. To overcome such challenges caused by ultra-high dimensionality, \cite{fan2008sure} proposed a sure independence screening (SIS) method, which aims to screen out the redundant features by ranking their marginal Pearson correlations. The SIS method is named after the sure independence screening property, which states the selected subset of features contains all the active ones with probability approaching one. The promising numerical performance soon made SIS popular among ultra-high dimensional studies \citep{liu2015selective, liu2020variable}.  The sure screening idea has been applied to many important statistical problems including generalized linear model \citep{fan2010sure}, multi-index semi-parametric model \citep{zhu2011model}, nonparametric model \citep{fan2011nonparametric, liu2014feature}, quantile regression \citep{he2013quantile, wu2015conditional} and compressed sensing \citep{xue2011sure} among others.

In addition to the sure screening property, we argue an appealing screening method should satisfy the following two properties. First, the screening method should be model-free in the sense that it can be implemented without specifying a regression model. In the ultra-high dimensional regime, it is challenging, if not impossible, to specify a correct regression model before removing a huge number of redundant features. Hence, the model-free property is desired as it guarantees the effectiveness of the screening method in the presence of model misspecification. The model-free screening method becomes a hot research topic in recent years \citep{zhu2011model, li2012feature, mai2015fused}.
The second property is data-adaptive which means the screening method should not be sensitive to assumptions like independence, sub-Gaussianity, and univariate response. Such assumptions are usually not satisfied under ultra-high dimensional settings. Even met on the population level, they can be violated in the observed sample due to ultra-high dimensionality. Therefore, the screening methods that are sensitive to such assumptions may perform poorly in real applications. The data-adaptive screening methods also draw a certain amount of attention recently. For instance, \cite{he2013quantile}, \cite{wu2015conditional} and \cite{ma2017variable}, among others, considered quantile-based screening which adapts to heavy-tailed data. In addition, \cite{wang2012factor} and \cite{fan2020factor} developed screening methods for strongly correlated features.

Unfortunately, none of the aforementioned screening methods enjoys sure screening, model-free and data-adaptive properties simultaneously. For example, the SIS is tailored to the linear regression and depends on the independence assumption. \cite{li2012feature} developed a model-free sure independence screening procedure based on the distance correlation. However, its sure screening property requires sub-exponential assumption for features and response. The Kolmogorov distance based screening method proposed in \cite{mai2012kolmogorov} is robust against heavy-tailed data, but it only works for binary classification problems.
\cite{pan2016ultrahigh} proposed a pairwise sure screening procedure for linear discriminant analysis, which is not sensitive to the tail behavior of the features, but requires balanced categories.

In this paper, we propose a model-free and data-adaptive feature screening method named PC-Screen. The PC-Screen is based on ranking the projection correlations between features and response variables.  The projection correlation, proposed by \cite{zhu2017projection}, is a measure of dependence between two random vectors which enjoys several nice probability properties. The PC-Screen does not require specifying any regression model and is insensitive to the correlations and moment conditions of the dataset. As the projection correlation is dimension-free to both random vectors, the PC-Screen can be applied to multi-task learning problems \citep{caruana1997multitask}. For instance, we can find a parsimonious set of features that are jointly dependent on the multivariate response.
As existing asymptotic results are not suitable for the  ``large $p$ small $n$ problem" studied in this paper, we also develop non-asymptotic concentration-type inequalities for empirical projection correlation. Further, we demonstrate the proposed PC-Screen enjoys not only the sure screening property but also a stronger result called rank consistency property. The only condition required is a minimum signal strength gap between active and inactive features. The extensive numerical examples show the proposed method wins the horse racing against its competitors in various scenarios.

Most feature screening methods depend on some threshold parameter that controls the cut-off between active and inactive features. The optimal choice of the threshold typically depends on unknown parameters. Under a specific model assumption, the threshold can be selected by cross-validation or information criteria approaches. However, when no model assumption is imposed, such parameter selection approaches are not applicable as the loss function which measures the goodness of fit is not well defined. In addition, existing screening methods tend to sacrifice the false discovery rate for sure screening property by choosing a conservative threshold parameter, leading to an inflated model size. Recently, \cite{zhao2012principled} studied the sure independence screening for Cox models with a principled method
to select the threshold aiming at controlling the false positive rate. \cite{song2014censored} proposed a censored rank independence screening for survival data and chose the threshold by estimating the proportion of active features. The validity of this procedure relies on the independence assumption of multiple test statistics which may easily get violated in the model-free setup.
In this paper, we tackle the issue of threshold selection with a two-step procedure named PC-Knockoff. In the first step, we apply the PC-Screen method to obtain an over-fitted subset of moderate size from the ultra-high dimensional features. In the second step, we construct knockoff counterparts for the features which survive in the first step. Conditioning on the sure screening of the first step, we further select a parsimonious model with FDR controlled below a pre-specified level with a statistic that utilizing the knockoff features. Theoretical analysis shows that when the pre-specified level is not too aggressive, the PC-Knockoff procedure enjoys sure screening property, as well as conditional FDR control simultaneously with high probability. We also validate the theoretical findings with various numerical examples.

The rest of the paper is organized as follows. In Section \ref{sec:screen}, we briefly review the definition and properties of projection correlation and demonstrate its non-asymptotic properties. Then we propose the PC-Screen procedure and show its sure screening and rank consistency properties under very mild conditions. Section \ref{sec:FDR} studies the PC-Knockoff procedure which selects the threshold
with FDR control as well as sure screening property.
Section \ref{sec:sim} assesses the finite sample performance of the
PC-Screen and PC-Knockoff methods with several simulated examples and a real
data application. We briefly summarize the paper in Section \ref{sec:con}.
Due to the limited space, we provide the proofs of some theoretical results
in the Appendix, and relegate the proofs of remaining results as well as
some additional numerical examples to the a supplementary file.

\section{Model-free and data-adaptive screening procedure}
\label{sec:screen}

\subsection{Projection correlation}
\label{sec:PC}

To pave the way for the proposed screening procedure, we first provide some background on the projection correlation and its properties introduced in \cite{zhu2017projection}.  Let $\bx \in \R^p$ and $\by \in \R^q$ be two random vectors. The projection correlation is elicited by the following independence testing problem,
$$
H_0: \bx \text{ and } \by \ \text{are independent} \quad \text{versus} \quad H_1: \text{ otherwise.}
$$
The null hypothesis holds if and only if $U=\balpha^{\T}\bx$ and $V=\bbeta^{\T}\by$ are independent for all unit vectors $\balpha$ and $\bbeta$. Let $F_{U,V}(u, v)$ be the joint distribution of $(U, \ V)$, and $F_U(u)$ and $F_V(v)$ be the marginal distributions of $U$ and $V$. The squared projection covariance is defined as
\begin{equation}\label{eq:pcov_1}
\begin{aligned}
   \pcov(\bx, \by)^2
   &=\iiint (F_{U,V}(u, v) - F_U(u)F_V(v))^2 \,d F_{U,V}(u, v) \,d \balpha \,d\bbeta \\
   &=\iiint \cov^2 \{I(\balpha^{\T}\bx\leq u), \ I(\bbeta^{\T}\by\leq v)\} \,d F_{U,V}(u, v) \,d \balpha \,d\bbeta, \\
\end{aligned}
\end{equation}
where $I(\cdot)$ is the indicator function.
Furthermore, the projection correlation between $\bx$ and $\by$ is defined as the square root of
\begin{equation}\label{eq:pc_1}
\pc(\bx, \by)^2  =\frac{\pcov(\bx, \by)^2 }{\pcov(\bx, \bx) \pcov(\by, \by) },
\end{equation}
and we follow the convention $0/0=0$.

In general $0\leq \pc(\bx, \by) \leq 1$, testing whether $\bx$ and $\by$ are independent amounts to testing whether $\pc(\bx, \by)=0$. The projection correlation is a measure of dependence between two random vectors and enjoys some appealing properties. Let $\bx$ and $\by$ be two random vectors with continuous marginal and joint probability distributions, $\pc(\bx, \by)=0$ if and only if $\bx$ and $\by$ are independent.
We remark here this property does not hold in general without the assumption that $(\bx,\by)$ is jointly continuous. When $\bx$ and $\by$ are two dependent discrete random variables that are constructed in a similar fashion as in \cite{hoeffding1948non}, it is possible that we have $\pc(\bx, \by)=0$.



\cite{zhu2017projection} gives an explicit formula for the squared projection covariance in (\ref{eq:pcov_1}).
Let $(\bx_1, \by_1),\ \ldots , \ (\bx_5, \by_5)$ be 5 independent random copies of $(\bx, \by)$, then
\begin{equation}
\begin{aligned}
  \pcov(\bx, \by)^2 =& S_1 + S_2 -2S_3 \\
 = & \E\left[\acos \Big\{\frac{(\bx_1-\bx_3)^{\T}(\bx_4-\bx_3)}{\|\bx_1-\bx_3\|\|\bx_4-\bx_3\|}\Big\}
\acos \Big\{\frac{(\by_1-\by_3)^{\T}(\by_4-\by_3)}{\|\by_1-\by_3\|\|\by_4-\by_3\|}
\Big\}\right] \\
+& \E\left[\acos \Big\{\frac{(\bx_1-\bx_3)^{\T}(\bx_4-\bx_3)}{\|\bx_1-\bx_3\|\|\bx_4-\bx_3\|}\Big\}
\acos \Big\{\frac{(\by_2-\by_3)^{\T}(\by_5-\by_3)}{\|\by_2-\by_3\|\|\by_5-\by_3\|}
\Big\}\right] \\
-2 & \E\left[\acos \Big\{\frac{(\bx_1-\bx_3)^{\T}(\bx_4-\bx_3)}{\|\bx_1-\bx_3\|\|\bx_4-\bx_3\|} \Big\}
\acos \Big\{\frac{(\by_2-\by_3)^{\T}(\by_4-\by_3)}{\|\by_2-\by_3\|\|\by_4-\by_3\|}
\Big\}\right],
\end{aligned}
\label{eq:pcov_2}
\end{equation}
where $\|\cdot\|$ is the $L_2$ norm. Equation (\ref{eq:pcov_2}) shows that the projection covariance only depends on the vectors through forms $(\bx_k-\bx_l)/\|\bx_k-\bx_l\|$ and $(\by_k-\by_l)/\|\by_k-\by_l\|$ whose second moments are unity. This gives us the intuition that the projection covariance is free of the moment conditions on $(\bx, \by)$ which are usually required by some other measurements, such as distance correlation \citep{li2012feature}.

Let $\bX=(\bx_1, \ \ldots \ , \bx_n)^{\T}$ and $\bY=(\by_1, \ \ldots \ , \by_n)^{\T}$ be an observed sample of $(\bx, \by)$.
Equation (\ref{eq:pcov_2}) leads to a straightforward estimator of $\pcov(\bx, \by)^2$ based on a $U$-statistic, yet it is difficult to calculate \citep{szekely2010brownian}. An equivalent form of the $U$-statistic is given in  \cite{zhu2017projection}. In particular, the squared sample projection variance and covariance between $\bX$ and $\bY$ can be calculated as
\begin{equation}
\begin{aligned}
\widehat{\pcov}(\bX, \bY)^2 & = n^{-3} \sum\limits_{k,l,r=1}^n A_{klr}B_{klr},
 \\
\widehat{\pcov}(\bX, \bX)^2 & = n^{-3} \sum\limits_{k,l,r=1}^n A_{klr}^2, \ \text{and} \
\widehat{\pcov}(\bY, \bY)^2 & = n^{-3} \sum\limits_{k,l,r=1}^n B_{klr}^2,
\end{aligned}
\label{eq:euqi_u_stat}
\end{equation}
where $k,l,r=1, \dots ,n, $
\begin{align*}
a_{klr} & = \arccos \left\{
\frac{(\bx_k-\bx_r)^{\T}(\bx_l-\bx_r)}
{\|\bx_k-\bx_r\|\|\bx_l-\bx_r\|} \right\},
\quad a_{klr}=0 \ \text{if} \ k=r \ \text{or} \ l=r,
\\
\bar{a}_{k \cdot r} & =n^{-1}\sum\limits_{l=1}^n a_{klr}, \quad
\bar{a}_{\cdot l r} =n^{-1}\sum\limits_{k=1}^n a_{klr}, \quad
\bar{a}_{\cdot \cdot r} =n^{-2}\sum\limits_{k=1}^n \sum\limits_{l=1}^n a_{klr},
\\
A_{klr} & =a_{klr}-\bar{a}_{k \cdot r}- \bar{a}_{\cdot l r}+ \bar{a}_{\cdot \cdot r},
\\
b_{klr} & = \arccos \left\{
\frac{(\by_k-\by_r)^{\T}(\by_l-\by_r)}
{\|\by_k-\by_r\|\|\by_l-\by_r\|} \right\}, \quad b_{klr}=0 \ \text{if} \ k=r \ \text{or} \ l=r,
\\
\bar{b}_{k \cdot r} & =n^{-1}\sum\limits_{l=1}^n b_{klr}, \quad
\bar{b}_{\cdot l r} =n^{-1}\sum\limits_{k=1}^n b_{klr}, \quad
\bar{b}_{\cdot \cdot r} =n^{-2}\sum\limits_{k=1}^n \sum\limits_{l=1}^n b_{klr},
\\
B_{klr} & =b_{klr}-\bar{b}_{k \cdot r}- \bar{b}_{\cdot l r}+ \bar{b}_{\cdot \cdot r}.
\end{align*}
Then the sample projection correlation between $\bX$ and $\bY$ is defined as the square root of
\begin{equation}
\widehat{\pc}(\bX, \bY)^2=
\frac{\widehat{\pcov}(\bX , \bY)^2}
{\widehat{\pcov}(\bX, \bX)\widehat{\pcov}(\bY, \bY)}.
\label{eq:samplePC}
\end{equation}
Based on (\ref{eq:euqi_u_stat}), the sample projection correlation can be computed in $O(n^3)$.

We first provide exponential-type deviation inequalities for squared sample projection covariance and correlation.

\begin{theorem}\label{thm:pc}
  For any $0 < \varepsilon < 1$ satisfying  $n \geq 10 \pi^2/ \varepsilon$, there exists positive constants $c_1$ and $c_2$, such that
$$
    \mathrm{Pr}\left\{ | \widehat{\pcov}(\bX, \bY)^2 - \mathrm{Pcov}(\bx, \by)^2 | > \varepsilon \right\} \leq c_1\exp\{-c_2n \varepsilon^2\}, \ \text{and} \
$$
  \begin{equation*}
    \mathrm{Pr}\left\{ | \widehat{\pc}(\bX, \bY)^2 - \mathrm{PC}(\bx, \by)^2 | > \varepsilon \right\} \leq 5c_1\exp\{-c_2 \sigma n\varepsilon^2\},
  \end{equation*}
    where $\sigma = \min\{ \sigma_x^3 \sigma_y^3/4M^4, \sigma_x^2 \sigma_y^2/4M^4 \}$, $\sigma_x = \pcov(\bx, \bx)^2$, $\sigma_y = \pcov(\by, \by)^2$ and $M = 4\pi^2$.

\label{error_bound}
\end{theorem}
The proof of Theorem \ref{error_bound} is based on an exponential-type deviation inequality for $U$-statistic and can be found in the supplementary material.
The above exponential inequalities do not depend on the dimensionality and moment conditions of both random vectors. The exception probability decays exponentially with sample size $n$ which guarantees good finite sample performance of the proposed estimator.

\subsection{PC-Screen procedure}\label{sec:pc_screen}

In this subsection, we propose a model-free and data-adaptive  screening
procedure utilizing the nice properties of projection correlation. Let
$\by=(Y_1, \ldots, Y_q)^{\T}$ be the response vector of $q$ variables and
$\bx=(X_1, \ldots, X_p)^{\T}$ be the vector of $p$ features. To avoid the
trivial discussion, we restrict ourselves to the non-degenerate case. That is,
$\min_{1 \leq k \leq p} \pcov(X_k, X_k)^2 \geq \sigma_0^2$ and $\min_{1 \leq k
\leq q}\pcov(Y_k, Y_k)^2 \geq \sigma_0^2$ for some $\sigma_0>0$.

Denote $F(\by|\bx)$ the conditional distribution function of $\by$ given $\bx$. Without specifying any regression model of $\by$ on $\bx$, we define the index set of active features by
\begin{equation*}
  \calA = \{ k: F(\by | \bx) \ \text{functionally depends on} \  X_k, k=1, \ldots, p \}.
\end{equation*}
The number of active features is $s=|\calA|$, the cardinality of $\calA$. The
features that do not belong to $\calA$ are inactive features. We use $\calA^c$,
the complement of $\calA$, to denote the index set of inactive features. The
above setting abstracts a large number of sparse regression problems including
linear model, generalized linear model, additive model, semi-parametric model,
non-linear model and so on. Moreover, it also allows the multivariate response
and grouped predictors.

Suppose that  $\{(\bx_i, \by_i), \ i=1, \ldots, n\}$ is a random sample of
$(\bx, \by)$. Denote $\bX=(\bx_1, \ldots, \bx_n)^{\T}$ and $\bY=(\by_1, \ldots,
\by_n)^{\T}$. In the ultra-high dimensional regime, it is natural to assume
that the number of features $p$ greatly exceeds the sample size $n$, but the
number of active features $s$ is smaller than $n$. For a given feature $X_k$,
$k=1, \ldots, p$, a sufficient condition for $X_k$ to be an inactive feature is
the independence between $X_k$ and $\by$. This, together with Theorem
\ref{error_bound}, motivates us to screen out the features whose projection
correlations with $\by$ are close to $0$. As a result, we estimate the set of
active features by
\begin{equation*}
  \hat{\calA}(\delta) = \{ k: \widehat{\pc}(\bX_k, \bY)^2 \geq \delta, 1 \leq k \leq p\},
\end{equation*}
where $\delta$ is a pre-specified positive threshold, and $\bX_k$ is the $k$th column of $\bX$.
With a proper choice of $\delta$, we show that the proposed feature screening procedure enjoys the sure screening property, which states that with probability approaching 1, all active features are included in $\hat\calA(\delta)$. We name this feature screening procedure as projection correlation based screening, or PC-Screen.
\medskip

Denote $\omega_k \equiv \pc(X_k, \by)^2$ and  $\hat\omega_k \equiv
\hat\pc(\bX_k, \bY)^2$ to be the squared population and sample projection
correlations between the $k$th feature and response, respectively. To analyze
the property of PC-Screen, we impose the following minimum signal strength
condition.

\noindent
{\bf Condition 1. (Minimum signal strength)}
\begin{itemize}
\item[(a)]For some $c_3>0$ and $0 \leq \kappa<1/2$,
$ \min_{k \in \calA}\omega_k \geq 2c_3 n^{-\kappa}.$
\item[(b)]For some $c_3 > 0$ and $0 \leq \kappa < 1/2$,
$\min_{k \in \calA} \omega_k - \max_{k \in \calA^c} \omega_k \geq 2c_3 n^{-\kappa}.$
\end{itemize}

\begin{remark}
Condition 1 (a) is a  minimum signal strength condition  that assumes the
squared projection correlations between active features and response are
uniformly bounded below, and cannot converge to zero too fast as $n$ diverges.
Condition 1 (b) imposes an assumption on the gap of signal strength between
active and inactive features. Condition 1 (a) is weaker than Condition 1 (b),
since $\omega_k$ is always non-negative. Such minimum signal strength condition
can be viewed as a sparsity assumption that guarantees active features to be
distinguished from the inactive ones. Condition 1 is a very mild condition as
we allow the minimum signal strength to converge to zero as the sample size
diverges. Empirically, one may verify Condition 1 by plotting the sorted
empirical projection correlations of features in descending order and visually
identifying whether there exists an ``elbow" shape, or whether the sequence can
be segmented into two groups. A more rigorous verification approach is to
consider a multiple-testing procedure. For each feature, one can calculate the
projection correlation based test statistic proposed in
\cite{zhu2017projection}. Since it is not the main focus of this paper, we do
not pursue more details.
\end{remark}

The following two theorems state the sure screening property and the rank consistency property of PC-Screen.

\begin{theorem}[Sure screening]\label{thm:ss}
  Under Condition 1 (a), take $\delta \leq \min_{k \in \calA}\omega_k/2$, we have
  \begin{equation}
    \mathrm{Pr}\Big(\calA \subseteq \hat{\calA}(\delta)\Big) \geq 1 - O\Big(s\exp\{ -c_4n^{1-2 \kappa}\}\Big),
  \end{equation}
  where $c_4$ is a positive constant.
\end{theorem}

In Theorem \ref{thm:ss}, if we set $\delta = c_3n^{-\kappa}$, which satisfies the condition $\delta \leq \min_{k \in \calA}\omega_k/2$, we have
\begin{equation}
  \mathrm{Pr}\Big(\calA \subseteq \hat{\calA}(c_3n^{-\kappa})\Big) \geq 1 - O(s\exp\{ -c_4n^{1 - 2 \kappa} \}).
  \label{remk:pcss}
\end{equation}
From (\ref{remk:pcss}), we know that if $\delta = c_3n^{-\kappa}$, all active features are selected with probability approaching 1 as $n \rightarrow \infty$. In fact, any choice of $\delta \leq c_3n^{- \kappa}$ leads to the sure screening property. With the same choice of $\delta$, \cite{li2012feature} showed that the distance correlation based screening method (DC-SIS) satisfies
\begin{equation*}
  \mathrm{Pr}\Big(\calA \subseteq \hat{\calA}(c_3n^{-\kappa})\Big) \geq 1 - O(s\exp\{ -c_4'n^{1 - 2 (\kappa + \eta )} \} + n \exp\{-c_4'' n^\eta\} ),
\end{equation*}
where $c_4'$, $c_4''$ and $\eta$ are positive constants.
Thus, PC-Screen achieves a faster rate than the DC-SIS since (1) we do not have the  term $n \exp\{-c_4'' n^\eta\}$ and (2) we do not have an extra $\eta$ in the power of the first term. The faster rate of PC-Screen is due to the fact that projection correlation is not sensitive to dimensionality and free of moment conditions.

\begin{theorem}[Rank consistency]
Under Condition 1 (b), we have
\begin{equation*}
 \mathrm{Pr}\left(\underset{k\in \calA}{\min}\ \homega_k - \underset{k\in \calA^c}{\max}\ \homega_k > 0 \right) > 1 - O(p\exp\{-c_5n^{1-2 \kappa}\}),
\end{equation*}
where $c_5$ is some positive constant. If $\log p = o(n^{1-2 \kappa})$ with $0 \leq \kappa < 1/2$, then we have
$$
  {\lim\inf}_{n\rightarrow\infty} \left(\underset{k\in \calA}{\min}\ \homega_k - \underset{k\in \calA^c}{\max}\ \homega_k \right) > 0, \ \text{almost surely}.
$$
\label{rank_consistency}
\end{theorem}
The rank consistency in Theorem \ref{rank_consistency} is a stronger result than the sure screening property. When the signal strength gap between active and inactive features satisfies Condition 1 (b), the active features are always ranked ahead of the inactive ones with high probability. In other words, there exists a choice of $\delta$ on the solution path that can perfectly separate the active and inactive sets with high probability.

\section{Screening with FDR control}\label{sec:FDR}

\subsection{Motivation}

In the PC-Screen procedure, the threshold $\delta$ controls the cut-off between active and inactive features. Theorem \ref{thm:ss} suggests choosing $\delta=cn^{-\kappa}$ for some positive constants $c$ and $\kappa <1/2$. With certain model assumptions, the threshold $\delta$ (or equivalently $c$ and $\kappa$) can be selected by cross-validation or information criterion approaches. However, in the model-free setup, such approaches are not directly applicable as the loss functions that measure the goodness of fit are not well defined.
More recently, \cite{zhao2012principled} and \cite{song2014censored} suggested to select the threshold by controlling the false positive rate and estimating the proportion of active features $s/p$, respectively. These two methods are tailored for survival analysis and are not directly applicable to model-free setup.

In practice, one can arbitrarily select a conservative threshold to ensure that all active features are included with high probability. However, it will include too many inactive features and inflate the FDR. Therefore, selecting the threshold parameter for model-free screening methods will inevitably cause the issue of balancing the trade-off between the sure screening property and FDR control.
In this section, we propose a two-step procedure to address this issue by utilizing knockoff features. The construction of knockoff features does not depend on the regression model, and hence is suitable for the model-free setting. Also, the second-order knockoff features can be easily computed by an equicorrelated construction or solving a semidefinite program \citep{barber2015controlling}. The proposed procedure enjoys sure screening property and controls FDR simultaneously with high probability.

\subsection{Knockoff features: a brief review}\label{sec:4.2}

Recently, knockoff has drawn huge attention due to its success in variable selection and many important applications such as genome-wide association studies. The concept of knockoff was first proposed in \cite{barber2015controlling} for the fixed design matrix problem and then extended to the random design matrix setting known as Model-X knockoffs \citep{candes2018panning}. For a more detailed development of knockoff, see \cite{barber2015controlling,candes2018panning,fan2019rank} and references therein. In this subsection, we briefly review the notations and definitions of knockoffs for further discussions.

Let $\by=(Y_1, \ldots, Y_q)^{\T} $ be a response vector of $q$ variables and $\bx=(X_1, \ldots, X_p)^{\T}$ be a covariate of $p$ features.
We say $\widetilde{\bx}=(\widetilde{X}_1, \ldots, \widetilde{X}_p)$ is a knockoff copy of $\bx$ if it satisfies the following conditions.

\noindent
{\bf Condition 2. (Exact knockoff features)}
\begin{itemize}
\item[(a)] Swap $X_j$ with its knockoff counterpart $\widetilde{X}_j$ does not change the joint distribution of $(\bx, \widetilde{\bx})$ for $j=1, \ldots, p$;

\item[(b)] Given $\bx$, $\widetilde{\bx}$ is independent of $\by$, i.e., $\widetilde{\bx} \independent \by | \bx$.
\end{itemize}
\vspace*{-0.1in}
\begin{remark}
Condition 2 (a) requires the original and knockoff features to be pairwise exchangeable. Condition 2 (b) indicates that knockoff features are conditionally independent of response variables; this is trivially satisfied if $\widetilde{\bx}$ is generated without using the information of $\by$.
\end{remark}

Constructing knockoff features that exactly follow Condition 2 is challenging, especially when the dimensionality of features $p$ is large. In general, generating exact knockoff features requires the knowledge of the underlying distribution of $\bx$, which is usually not available in practice. \cite{barber2015controlling} studied the variable selection problem with knockoffs for fixed design matrix $\bX \in \RR^{n\times p}$. The construction of exact knockoffs is not necessary if assuming the response $\by$ follows a linear regression model with Gaussian error and $n \geq 2p$.
A more recent study \citep{candes2018panning} proposed to construct exact knockoff features in a manner of generating sequential conditional independent pairs under the assumption that the distribution of $\bx$ is known.

Without the knowledge of the distribution of $\bx$, one can construct approximate second-order knockoff features such that $(\bx, \widetilde{\bx})$ is pairwise exchangeable with respect to the first two moments. In other words, the mean vector and covariance matrix of $(\bx,\tilde\bx)$ is invariant if we swap $X_j$ and $\widetilde{X}_j$ for any $j=1, \ldots, p$. The invariant of mean is trivial, and can be achieved by forcing $\E(\bx)=\E(\widetilde{\bx})$.
Suppose $\cov(\bx)= \bSigma$, the second-order pairwise exchangeable condition is equivalent to
\begin{equation}
\cov(\bx, \widetilde{\bx})=\bG, \quad \text{where}  \quad
\bG=\left[\begin{matrix}
\bSigma & \bSigma-\mathrm{diag}\{ \bh \} \\
\bSigma-\mathrm{diag}\{ \bh \} & \bSigma
\end{matrix}\right],
\label{eq:G}
\end{equation}
where $\bh = (h_1, \ldots, h_p)^\T$ is a vector that makes $\bG$ a positive semidefinite covariance matrix.

\cite{barber2015controlling} introduced two approaches to construct the second-order knockoffs. The first approach is known as the {\it equicorrelated construction}, which sets
\begin{equation}
      h_j = 2 \lambda_{\min}(\bSigma)\vee 1 \ \ \text{for} \  j=1, \  \ldots, \  p,
\label{eq:equi}
\end{equation}
where $\lambda_{\min}(\bSigma)$ is the minimum eigenvalue of $\bSigma$, and $a \vee b$ denotes the larger one between $a$ and $b$. The second method, named {\it semidefinite programme}, finds $h_j$ by solving a semidefinite program of the following form
\begin{equation}
\begin{aligned}
      \text{minimize} \quad &\sum_{j}|1-h_j|, \\
      \text{subject to} \quad &h_j \geq 0, \text{diag}\{\bh\} \preceq 2\bSigma.
\end{aligned}
\label{eq:sdp}
\end{equation}
However, both methods are not directly applicable in the high-dimensional scenarios since both (\ref{eq:equi}) and (\ref{eq:sdp}) require $2p < n$.

\begin{remark}
When $(\bx, \widetilde{\bx})$ is jointly Gaussian, the equivalence of the first two moments implies the equivalence of the joint distribution and hence (\ref{eq:G}) constructs exact knockoff features. However, when the Gaussian assumption does not hold, the accuracy of the second-order approximation depends on the impact of ignoring higher-order moments of $\bx$.
Also, the covariance matrix $\bSigma$ is usually unknown and needs to be estimated. Hence the estimation accuracy of $\bSigma$ also affects the validation of the second-order approximation of exact knockoff features. In addition to sample covariance estimator, more sophisticated estimators of $\bSigma$ can be obtained under additional structure or moment conditions, see \cite{ fan2013large} and \cite{ke2018user}, among others. As the above two issues are not of key interest in this paper, we do not pursue further in these directions.
\end{remark}

\subsection{FDR control with knockoff features}\label{sec:4.3}
Suppose $\widetilde{\bx}=(\widetilde{X}_1, \ldots, \widetilde{X}_p)^\T$ is a knockoff copy of $\bx=(X_1, \ldots, X_p)^{\T}$, we propose to measure the population level dependence between $X_j$ and response $\by$ by the following quantity
\begin{equation}\label{eq:wj}
W_j={\pc}(X_j, \by){^2} - {\pc}(\widetilde{X}_j, \by){^2}, \quad j=1, \ldots, p,
\end{equation}
where ${\pc}(X_j, \by)$ and ${\pc}(\widetilde{X}_j, \by)$ are the projection correlations as defined in \eqref{eq:pc_1}. When $\widetilde{X}_j$ is an exact knockoff feature of $X_j$, $W_j$ will be a non-negative quantity. Further, $W_j>0$ implies the distribution of $\by$ depends on $X_j$ and $W_j = 0$ if $\by$ is independent with $X_j$ conditioning on active features.

Given a random sample $\{\bY, \bX, \widetilde{\bX}\}$ drawn from $\{\by, \bx, \widetilde{\bx}\}$, we estimate ${W}_j$ by
\begin{equation}
\widehat{W}_j=\widehat{\pc}(\bX_j, \bY){^2} - \widehat{\pc}(\widetilde{\bX}_j, \bY){^2}, \quad j=1, \ldots, p.
\label{eq:hatwj}
\end{equation}
where $\widehat{\pc}(\bX_j, \bY)$ and $\widehat{\pc}(\widetilde{\bX}_j, \bY)$ are sample projection correlations defined in \eqref{eq:samplePC}. Intuitively, a large positive value of $\hat{W}_j$ provides some evidence that the distribution of $\by$ depends on $X_j$. On the other hand, if $X_j$ is an inactive feature, $|\hat{W}_j|$ is likely be be small and $\hat{W}_j$ is equally likely to be positive or negative.
This intuition is justified by the following lemma.
\begin{lemma}
      Let $\tilde\bx$ be an exact knockoff copy of $\bx$ and $\calA^c =\{j_1, \dots, j_r\}$. Then
      \begin{enumerate}[(i)]
            \item $W_{j_k} = 0$ for all $j_k \in \calA^c$.
            \item Conditioning on $|\hat\bw| = (|\hat W_1|, \dots, |\hat W_p|)^\T$, $I_{j_1}, \dots, I_{j_r}$ follow i.i.d. $Bernoulli(0.5)$, where $I_{j_k} = 1$ if $\hat W_{j_k} > 0$ and 0 otherwise.
      \end{enumerate}
\label{lem:null_wj}
\end{lemma}

For a fixed threshold $t>0$, the false discovery proportion (FDP) is defined as
\begin{equation*}
{\FDP}(t) = \frac{\#\{j \in \calA^c: \hat{W}_j \geq t \}}{\#\{j: \hat{W}_j \geq t \}},
\end{equation*}
where $\#\{\cdot\}$ is the cardinality of a set and we follow the convention that $0/0=0$.
The false discovery rate (FDR) is defined as the expectation of FDP, i.e., $\mathrm{FDR}(t) = \E\left[{\FDP}(t)\right]$.
According to Lemma \ref{lem:null_wj}, $\hat{W}_j$ is equally likely to be positive or negative if $X_j$ is an inactive feature. Therefore, we have
\begin{equation*}
\#\{j \in \calA^c: \hat{W}_j \geq t \} \approx \#\{j \in \calA^c: \hat{W}_j \leq -t \} \leq \#\{j: \hat{W}_j \leq -t \},
\end{equation*}
which leads to a conservative estimation of ${\FDP}(t)$,
\begin{equation*}
\widehat{\FDP}(t) = \frac{\#\{j: \hat{W}_j \leq -t \}}{\#\{j: \hat{W}_j \geq t \}}.
\end{equation*}

To control FDR at a pre-specified level $\alpha$, we follow the knockoff+ procedure
\citep{barber2015controlling} to choose the threshold $T_\alpha$ as
\begin{equation}\label{eq:ta}
T_{\alpha}= \min \left\{t \in \calW: \ \frac{1+\#\{j: \hat{W}_j \leq -t \}}{\#\{j: \hat{W}_j \geq t \}} \leq \alpha  \right\},
\end{equation}
where $\calW = \{|\hat{W}_j|: 1 \leq j \leq p\} / \{0\}$.
The extra term $1$ in the numerator makes the choice of $T_\alpha$ slightly more conservative.
Then, the active set is selected as
\begin{equation}\label{eq:ahat_ta}
  \hat{\calA}(T_\alpha) = \{ j: \hat{W}_j  \geq T_\alpha, 1 \leq j \leq p\}.
\end{equation}

\subsection{PC-Knockoff procedure}\label{sec:4.4}

The knockoff feature construction methods discussed in Section \ref{sec:4.2} require $2p < n$ and hence are not applicable to high-dimensional scenarios. To address this issue, we propose a two-step procedure, named PC-Knockoff, to screen active features and control the FDR. To avoid mathematical challenges caused by the reuse of sample, we follow the simple sample splitting idea,
which has been widely used in statistics and recent examples include hypothesis testing \citep[e.g.][]{Fan2019farmtest}, error variance estimation \citep[e.g.][]{chen2018error}, variable selection \citep[e.g.][]{barber2019knockoff}, large scale inference \citep[e.g.][]{fan2019rank}, and so on.
We partition the full sample into two disjoint subsamples with sample sizes $n_1$ and $n_2=n-n_1$.
More specifically, let $\bX^{(1)}\in\RR^{n_1\times p}$ and $\bX^{(2)}\in\RR^{n_2\times p}$ be a random partition of $\bX$, and let  $\bY$ follow the same partition. Without loss of generality, we write
\begin{equation*}
\bX =
\begin{bmatrix}
\bX^{(1)} \\
\bX^{(2)} \\
\end{bmatrix}
\quad \text{and} \quad
\bY =
\begin{bmatrix}
\bY^{(1)} \\
\bY^{(2)} \\
\end{bmatrix}.
\end{equation*}

\noindent The two steps of PC-Knockoff procedure are introduced as follows:

\noindent{\sc{(1) Screening step}:} We rank all $p$ features in descending order based on the sample projection correlation $\widehat\pc(\bX_j^{(1)}, \bY^{(1)})$. Then, we select the top $d$ features such that $2d < n_2$. Denote the set of selected $d$ features by $\widehat\calA_1$. In practice, one can set $d$ to be a relatively large value as long as it satisfies $2d<n_2$.

\medskip
\noindent{\sc{(2) Knockoff step}:}  Let
$$\bX^{(2)} = \left( \bX^{(2)}_{\hat\calA_1}, \bX^{(2)}_{\hat\calA^c_1} \right).$$
We construct knockoff features for $\bX^{(2)}_{\hat\calA_1}$ by either the {\it equicorrelated construction} as in \eqref{eq:equi}  or the {\it semidefinite programme} as in \eqref{eq:sdp}. Denoted $\widetilde\bX^{(2)}_{\hat\calA_1}$ the constructed knockoff features for $\bX^{(2)}_{\hat\calA_1}$. Then, we calculate
\begin{equation}
\widehat{W}_j=\widehat{\pc}(\bX^{(2)}_{\hat\calA_1, j}, \bY^{(2)}){^2} - \widehat{\pc}(\widetilde{\bX}^{(2)}_{\hat\calA_1,j}, \bY^{(2)}){^2}, \quad j=1, \ldots, d,
\end{equation}
where $\bX^{(2)}_{\hat\calA_1, j}$ and $\widetilde{\bX}^{(2)}_{\hat\calA_1,j}$ are the $j$th columns of $\bX^{(2)}_{\hat\calA_1}$ and $\widetilde\bX^{(2)}_{\hat\calA_1}$, respectively.
For a pre-specified FDR level $\alpha$, we use \eqref{eq:ta} to choose the threshold $T_\alpha$ and the set of selected active features is given by
\begin{equation*}
  \hat\calA(T_\alpha) = \{j: j\in\hat\calA_1, {\widehat W_j} \geq T_\alpha\}.
\end{equation*}
\medskip

The proposed two-step approach has two advantages. First, we apply the PC-Screen procedure to reduce the number of features from $p$ to $d$, which allows us to construct second-order knockoff features. Second, by ruling out $p-d$ inactive features in the screening step, we reduce the total computation cost from $O(n^3p)$ to $O(n_1^3p+n_2^3d)$. We summarize the PC-Knockoff procedure in Algorithm \ref{alg:fs_knockoff}. In fact, the Algorithm \ref{alg:fs_knockoff} provides a general framework for feature screening with FDR control. One can easily modify Algorithm \ref{alg:fs_knockoff} by replacing PC with other measurement statistics such as Pearson correlation, distance correlation, etc.

\begin{algorithm}
\caption{PC-Knockoff}
\label{alg:fs_knockoff}
\begin{algorithmic}
\STATE \textbf{Input}: $(\bX, \bY) \in \RR^{n\times p}\times \RR^{n\times q}$, $\alpha$,  $n_1$, and $d < {n_2}/{2}$, where $n_2=n-n_1$.\\
Partition $(\bX, \bY)$ into $(\bX^{(1)}, \bY^{(1)}) \in \RR^{n_1\times p}\times \RR^{n_1\times q}$ and $(\bX^{(2)}, \bY^{(2)}) \in \RR^{n_2\times p}\times \RR^{n_2 \times q}$.
\STATE \textbf{1. Screening step} \\
\STATE \quad \ For $j=1, \dots, p$, compute the sample squared PC: $\widehat \omega^{(1)}_j = {\widehat\pc(\bX_j^{(1)}, \bY^{(1)})^2}$. \\
       \quad \ Select the top $d$ features, i.e., $\hat\calA_1 = \{ j: \widehat \omega^{(1)}_j \ \text{is among the largest $d$} \}$.
\STATE \textbf{2. Knockoff step} \\
\STATE \quad \ Construct second-order knockoff features $\widetilde\bX^{(2)}_{\hat\calA_1}$ for $\bX^{(2)}_{\hat\calA_1}$ using \eqref{eq:equi} or \eqref{eq:sdp}.
\STATE \quad \ For all $j \in \hat\calA_1$, compute  $\widehat{W}_j = {\widehat\pc}(\bX_{\hat\calA_1,j}^{(2)}, \bY^{(2)}){^2} - {\widehat\pc}(\widetilde\bX_{\hat\calA_1,j}^{(2)}, \bY^{(2)}){^2}$.
\STATE \quad \ Choose the threshold $T_\alpha$ by solving (\ref{eq:ta}).
\STATE \textbf{Output}: Set of selected active features $\widehat\calA(T_\alpha) = \{j: j\in\hat\calA_1, {\widehat W_j} \geq T_\alpha\}$.
\end{algorithmic}
\end{algorithm}

\begin{remark} We want to add a note that Algorithm \ref{alg:fs_knockoff} is not tuning free as one needs to specify the subsample size $n_1$ and target dimension $d$ for the screening step. However, Algorithm \ref{alg:fs_knockoff} is not sensitive to the choice of these two hyper-parameters as Theorem \ref{thm:ss} guarantees the sure screening property of the screening step under mild conditions.
In practice, we suggest small $n_1$. As a result, we leave a relatively large subsample for the knockoff step. A relatively large $n_2$ allows more features to be selected in the screening step (larger $d$) and more accurate second-order knockoff features constructed in the knockoff step.
\end{remark}

\medskip

Denote $\calE$ the event that the sure screening property is satisfied in the screening step, i.e.,
\[
      \calE = \{ \text{All active features are selected in the screening step} \}.
\]
The sure screening property in Theorem \ref{thm:ss} ensures that, with a relatively large choice of $d$, the event $\calE$ holds with high probability. To be specific, let $\hat \omega^{(1)}_{(1)} \geq \hat \omega^{(1)}_{(2)} \geq \dots \geq \hat \omega^{(1)}_{(p)}$ be the order statistics of sample squared projection correlations based on $(\bX^{(1)}, \bY^{(1)})$. If Condition 1 (a) holds and $\hat \omega^{(1)}_{(d)} \leq c_3n_1^{-\kappa}$, then event $\calE$ holds with probability at least $1 - O(s\exp\{c_4n_1^{1-2 \kappa}\})$. Conditioning on $\cal E$, the following theorem states that the PC-Knockoff procedure can control the FDR of selected features under the pre-specified level of $\alpha$.

\begin{theorem}
Let $\wt \bX_{\hat\calA_1}$ be a knockoff copy of $\bX_{\hat\calA_1}$ satisfying Condition 2.
For any $\alpha \in [0,1]$, the set of selected features $\hat\calA(T_\alpha)$ given by Algorithm \ref{alg:fs_knockoff} satisfies
\[
      \mathrm{FDR} = \mathrm{E}\left[ \frac{\#\{ j:j\in \calA^c \cap \hat\calA(T_\alpha) \}}{\#\{ j:j\in\hat\calA(T_\alpha) \} \vee 1} \ \big| \ {\cal E}\right] \leq \alpha.
\]
\label{thm:pc_fdr}
\end{theorem}

Theorem \ref{thm:pc_fdr} states that, with exact knockoff features and conditioning on $\cal E$, the PC-Knockoff procedure can control the FDR under the pre-specified level $\alpha \in [0,1]$. As $\cal E$ occurs with probability close to 1, the $\mathrm{FDR}$ can be controlled even without conditioning on $\cal E$ \citep{barber2019knockoff}. We refer to \cite{barber2019knockoff} and \cite{fan2019rank} for more discussion regarding $\mathrm{FDR}$ control in two-step procedures.

\medskip

We hope the PC-Knockoff procedure can maintain sure screening property and control FDR simultaneously. This task is challenging as the procedure needs to balance the trade-off between type I and type II errors. To guarantee the sure screening property, the procedure is likely to select an over-fitted model which leads to higher type I errors as well as FDR. On the other hand, the FDR control forces a parsimonious but possibly under-fitted model that can increase type II errors and ruin the sure screening property. In the following, we study the conditions under which this challenging task can be achieved by the PC-Knockoff procedure.
Conditioning on $\calE$, the following theorem states that the simultaneous achievement of sure screening property and FDR control under level $\alpha$, which depends on the relationship between $s$ and $\alpha$.

\begin{theorem}\label{Thm_12}
Under the conditions of Theorem~\ref{thm:pc_fdr} and further assume $\min_{k \in \calA}W_j \geq 4c_3 n_2^{-\kappa}$ for some $c_3>0$ and $0 \leq \kappa < 1/2$.
      \begin{enumerate}[(i)]
            \item If $\alpha \geq 1/s $, we have $\text{Pr}(\calA \subseteq \hat\calA(T_\alpha)|\calE) \geq 1 - O(n_2 \exp\{-c_4 n_2^{1-2 \kappa} \})$.
            \item If $\alpha < 1/s$, we have $\text{Pr}(\calA \subseteq \hat\calA(T_\alpha)\cup\hat\calA(T_\alpha) = \varnothing|\calE) \geq 1 - O(n_2 \exp\{-c_4 n_2^{1-2 \kappa} \})$. Furthermore, if $s > 2$, we have $\text{Pr}(\calA \subseteq \hat\calA(T_\alpha)|\calE) \leq C(s) + O(n_2 \exp\{-c_4 n_2^{1-2 \kappa} \})$, where $0 <C(s)<1$ is a constant that only depends on $s$.
      \end{enumerate}
\label{thm:ss_fdr}
\end{theorem}

The part (i) of Theorem \ref{thm:ss_fdr} together with Theorem \ref{thm:pc_fdr}
suggest that if $\alpha$ is chosen to be greater or equal to $1/s$, the
PC-Knockoff procedure enjoys the sure screening property and controls FDR under
$\alpha$ with high probability. When $\alpha$ is chosen to be smaller than
$1/s$, we either recover the active set or end up with an empty set with high
probability. The probability of recovering the active set is upper bounded by
some constant $C(s)$ depending on $s$. Therefore there is no guarantee that
PC-Knockoff can select all active features while controlling FDR under
$\alpha$. As we know, the value of $\alpha$ controls the amount of type I
errors that we can tolerate. One can imagine that the smaller the $\alpha$ is,
the more challenging the task is to achieve sure screening and FDR control
simultaneously. It is because we allow fewer and fewer inactive features to be
selected. There is a phase transition that happens when $\alpha$ goes below
$1/s$. In order to satisfy the sure screening property, the procedure will fail
to control the FDR at $\alpha$ with non-negligible probability  if any inactive
feature is selected. In other words, the procedure has to exactly recover the
true active set to satisfy FDR control and sure screening simultaneously. We
numerically validate this phase transition phenomenon through a simulated
example in the supplementary file.

Theorem  \ref{thm:ss_fdr} discourages us to pursue a too aggressive $\alpha$ in practice as the PC-Knockoff procedure may lose the sure screening property. The phase transition between part (i) and part (ii) can also be used as a rule of thumb guideline to estimate $s$ in practice. For a sequence of grid points of $\alpha$ in $(0,1)$, we find the largest grid point $\alpha^*$ such that the PC-Knockoff procedure selects an empty set. Then, we can roughly estimate $\widehat{s}$ as the integer part of  $1/\alpha^*$.

\begin{remark}
As a two-step procedure, the power of the PC-Knockoff is conditioned on the event that the sure screening property is satisfied in the screening step, i.e., $\calE$.
This together with the results in Theorem \ref{thm:ss_fdr} yield that the probability that the PC-Knockoff procedure does not lose any power is at least $1 - O(s\exp\{-c_4n_1^{1-2 \kappa}\})-O(n_2 \exp\{-c_4 n_2^{1-2 \kappa} \})$. Empirically, one may follow the ``data recycling " idea proposed in \cite{barber2019knockoff} to improve the power of the PC-Knockoff procedure.
\end{remark}

As a byproduct of the PC-Knockoff procedure, the statistic $\hat W_j$ defined in \eqref{eq:hatwj}
can also be used to measure the marginal dependence between the response and the $j$th feature. Given a threshold $\delta$, we can screen a model based on $\hat W_j$ by considering the following set
\[
\hat\calA_W(\delta) = \{ j: \widehat W_j \geq \delta, 1 \leq j \leq p  \}.
\]
The next theorem states that the screening procedure based on $\hat W_j$ also enjoys the sure screening and rank consistency properties.

\begin{theorem}
Suppose $\widetilde \bX$ is an exact knockoff copy for $\bX$ and $\min_{k \in \calA} W_j \geq 2 c_3n^{-\kappa}$ for some positive constants $c_3>0$ and $0<\kappa<1/2$, then
\begin{enumerate}[(i)]
      \item $\mathrm{Pr}\Big(\calA \subseteq \hat\calA_W(\delta)\Big) \geq 1 - O\Big(s\exp\{ -c_4n^{1-2 \kappa} \}\Big)$ for $\delta \leq c_3n^{-\kappa}$.
      \item $\mathrm{Pr}\left(\underset{k\in \calA}{\min}\ \hat W_k - \underset{k\in \calA^c}{\max}\ \hat W_k > 0 \right) \geq 1 - O(p\exp\{-c_4n^{1-2 \kappa}\})$.
\end{enumerate}
\label{thm:byproduct}
\end{theorem}

Recall that the rank consistency property in Theorem \ref{rank_consistency} requires a minimum signal gap between active and inactive sets, that is
$\min_{k\in \calA} \omega_k - \max_{k \in \calA^c}{\omega_k} > 2c_3n^{-\kappa}$. However, the rank consistency result in Theorem \ref{thm:byproduct} only requires a minimum signal strength of active features and no condition is imposed on the inactive set. Due to the construction of $W_j$, signals of inactive features are canceled out by their knockoff counterparts. As a result, the rank consistency property holds even when some inactive features are spuriously correlated with the response. If we can construct high-quality knockoff features, the screening procedure based on $\hat W_j$ can be more powerful than PC-Screen.

\section{Numerical examples}\label{sec:sim}

\subsection{Screening performance}\label{sec:sim:part1}

In this subsection, we use simulated examples to assess the finite sample performance of the proposed projection correlation based feature screening procedure (PC-Screen) and compare it with sure independence screening \citep[SIS]{fan2008sure}, distance correlation based screening \citep[DC-SIS]{li2012feature} and bias-corrected distance correlation based screening \citep[bcDC-SIS]{szekely2014partial}.
Within each replication, we rank the features in descending order by the above four screening criteria and record the minimum model size that contains all active features. The screening performance is measured by the $5\%, 25\%, 50\%, 75\%$ and $ 95\%$ quantiles of the minimum model size over 200 replications.
Throughout this subsection, we denote $\bSigma = (\sigma_{ij})_{p\times p}$ with $\sigma_{ij} = 0.5^{|i-j|}$. To mimic ultra-high dimensional scenario, we set $n=100$ and $p=5,000 , 10,000$ for each example.


\subsubsection*{Example 1: Linear and Poisson models}

Consider the linear model
$
  Y = \bx^{\T}\bbeta + \varepsilon
$
with $\bbeta = (\bfm 1_5^{\T}, \bfm 0_{p-5}^{\T})^{\T}$. We generate covariates $\bx$ and $\varepsilon$ independently from the following 4 scenarios.

\begin{description}
  \item[\quad \quad Model 1.a:] \ $\bx \sim N (\bfm 0, \bfm\Sigma)$ and $\varepsilon \sim N (0, 1)$.
  \item[\quad \quad Model 1.b:] \ $\bx \sim N (\bfm 0, \bfm\Sigma)$ and $\varepsilon \sim \text{Cauchy}(0, 1)$.
  \item[\quad \quad Model 1.c:] \ $\bu \sim \text{Cauchy}(\bfm 0, \bI_p)$, $\bx = \bfm\Sigma^{1/2}\bu$ and $\varepsilon \sim N(0, 1)$.
  \item[\quad \quad Model 1.d:] \ $\bu \sim \text{Cauchy}(\bfm 0, \bI_p)$, $\bx = \bfm\Sigma^{1/2}\bu$ and $\varepsilon \sim \text{Cauchy}(0, 1)$.
\end{description}
In above models, $\text{Cauchy}(\bfm 0, \bI_p)$ stands for the $p$-dimensional standard Cauchy distribution which is heavy-tailed.  Hence, in Models 1.b -- 1.d, at least one of $\bx$ and $\varepsilon$ is  heavy-tailed.
We also consider the following two Poisson regression models
\begin{description}
  \item[\quad \quad Model 1.e:] \ (Continuous) $Y = \exp\{\bx^{\T}\bbeta\} + \varepsilon$, where $\varepsilon \sim N(0,1)$.
  \item[\quad \quad Model 1.f:] \ (Discrete) $Y \sim \text{Poisson}(\exp\{\bx^{\T}\bbeta\})$.
\end{description}
Let $\bbeta = (\bfm 2_5^{\T}, \bfm 0_{p-5}^{\T})^{\T}$ and we draw $\bx$ from $ N(\bfm 0, \bfm\Sigma)$. Model 1.e is the Poisson regression model with continuous response while the response in Model 1.f is discrete.

\medskip

The quantiles of the minimum model size that includes all 5 active features are summarized in Table \ref{tab_linear_reg}. In the linear benchmark Model 1.a, all four competitors perform well. For Models 1.b -- 1.d, the SIS completely fails at the presence of heavy-tailed features and errors. Both distance correlation-based methods struggle to maintain a reasonable model size at $75\%$ and $95\%$ quantiles. In contrast, PC-Screen works reasonably well in all scenarios and outperforms the other three methods by big margins.
Similarly, for the Poisson models 1.e and 1.f, PC-Screen can recover the true active set with a model size close to 5 while the other three methods can perform as bad as random guesses at $75\%$ and $95\%$ quantiles.

\begin{table}[htp]
\centering
\caption{The quantiles of minimum model size for linear and Poisson models in Example 1 over 200 replications. The true model size is 5.}
\label{tab_linear_reg}
\resizebox{1\textwidth}{!}{
\begin{tabular}{@{}rrrrrrrrrrrrrrr@{}}
\toprule
          && \multicolumn{5}{c}{\textbf{Model 1.a}}  & \phantom{ab} &\multicolumn{5}{c}{\textbf{Model 1.b}}  \\
          \cmidrule{3-7} \cmidrule{9-13}
          && 5\%       & 25\%   & 50\%   & 75\%   & 95\%   && 5\%      & 25\%   & 50\%   & 75\%   & 95\%   \\ \midrule
$p=5000$  &&           &      &       &       &        &&           &       &        &        &        \\
PC-Screen && 5.0       & 5.0  & 5.0   & 5.0   & 8.0    && 5.0       & 5.0   & 6.0    & 14.0   & 125.0  \\
DC-SIS    && 5.0       & 5.0  & 5.0   & 5.0   & 6.0    && 5.0       & 5.0   & 10.0   & 81.2   & 1305.0 \\
bcDC-SIS  && 5.0       & 5.0  & 5.0   & 5.0   & 6.0    && 5.0       & 5.0   & 6.0    & 20.5   & 231.4  \\
SIS       && 5.0       & 5.0  & 5.0   & 5.0   & 5.0    && 6.0       & 238.0 & 1833.0 & 3878.5 & 4915.0 \\
\midrule
$p=10000$ &&           &      &       &       &        &&           &       &        &        &        \\
PC-Screen && 5.0       & 5.0  & 5.0   & 5.0   & 7.0    && 5.0       & 5.0   & 8.0    & 23.0   & 233.5  \\
DC-SIS    && 5.0       & 5.0  & 5.0   & 5.0   & 6.0    && 5.0       & 6.8   & 21.0   & 204.2  & 3511.2 \\
bcDC-SIS  && 5.0       & 5.0  & 5.0   & 5.0   & 6.0    && 5.0       & 5.0   & 10.0   & 43.2   & 718.8  \\
SIS       && 5.0       & 5.0  & 5.0   & 5.0   & 5.0    && 16.0      & 703.2 & 3418.5 & 7432.0 & 9651.0 \\
\midrule
          && \multicolumn{5}{c}{\textbf{Model 1.c}}  & \phantom{ab} &\multicolumn{5}{c}{\textbf{Model 1.d}}  \\
          \cmidrule{3-7} \cmidrule{9-13}
          && 5\%       & 25\%   & 50\%   & 75\%   & 95\%   && 5\%      & 25\%   & 50\%   & 75\%   & 95\%   \\ \midrule
$p=5000$  &&           &      &       &       &        &&           &       &        &        &        \\
PC-Screen && 5.0       & 5.0  & 6.0   & 8.0   & 50.9   && 5.0       & 5.0   & 6.0    & 10.2   & 139.9  \\
DC-SIS    && 5.0       & 8.0  & 42.5  & 143.0 & 722.5  && 5.0       & 16.0  & 54.0   & 189.0  & 701.6  \\
bcDC-SIS  && 5.0       & 5.0  & 6.0   & 14.2  & 80.4   && 5.0       & 5.0   & 8.0    & 21.8   & 156.8  \\
SIS       && 5.0       & 39.0 & 81.5  & 374.0 & 3244.4 && 5.0       & 45.8  & 130.5  & 523.5  & 3241.0  \\
\midrule
$p=10000$ &&           &      &       &       &        &&           &       &        &        &        \\
PC-Screen && 5.0       & 5.0  & 6.0   & 8.0   & 92.4   && 5.0       & 6.0   & 6.0    & 13.0   & 176.2  \\
DC-SIS    && 5.0       & 18.2 & 78.0  & 277.5 & 1567.5 && 5.0       & 30.8  & 113.0  & 410.0  & 2036.5 \\
bcDC-SIS  && 5.0       & 5.0  & 7.0   & 19.2  & 179.1  && 5.0       & 5.0   & 10.0   & 27.0   & 412.6  \\
SIS       && 6.0       & 58.8 & 180.5 & 773.2 & 4189.0 && 8.9       & 73.0  & 244.5  & 959.8  & 5777.7 \\
\midrule
          && \multicolumn{5}{c}{\textbf{Model 1.e}}  & \phantom{ab} &\multicolumn{5}{c}{\textbf{Model 1.f}}  \\
          \cmidrule{3-7} \cmidrule{9-13}
          && 5\%       & 25\%   & 50\%   & 75\%   & 95\%   && 5\%       & 25\%   & 50\%   & 75\%   & 95\%   \\ \midrule
$p=5000$  &&           &        &        &        &        &&           &        &        &        &        \\
PC-Screen && 5.0       & 5.0    & 5.0    & 5.0    & 17.2   && 5.0       & 5.0    & 5.0    & 5.0    & 7.0    \\
DC-SIS    && 90.3      & 396.0  & 897.5  & 1762.5 & 3787.3 && 76.3      & 396.0  & 893.5  & 1764.8 & 3471.8 \\
bcDC-SIS  && 22.7      & 82.2   & 259.5  & 669.2  & 2358.7 && 15.0      & 87.2   & 266.5  & 821.2  & 2471.5 \\
SIS       && 178.6     & 604.8  & 1137.0 & 2319.8 & 4303.4 && 186.0     & 606.2  & 1210.5 & 2253.0 & 4261.6 \\
\midrule
$p=10000$ &&           &        &        &        &        &&           &        &        &        &        \\
PC-Screen && 5.0       & 5.0    & 5.0    & 6.0    & 23.0   && 5.0       & 5.0    & 5.0    & 5.0    & 12.0   \\
DC-SIS    && 138.0     & 788.8  & 1878.5 & 3301.0 & 6831.9 && 154.2     & 729.2  & 1725.5 & 3424.0 & 6832.8 \\
bcDC-SIS  && 45.7      & 163.5  & 534.5  & 1520.5 & 5415.2 && 30.0      & 175.8  & 509.0  & 1513.2 & 5580.0 \\
SIS       && 462.8     & 1276.8 & 2460.0 & 4164.5 & 8622.5 && 512.6     & 1271.2 & 2484.5 & 4281.0 & 8416.3 \\
\bottomrule
\end{tabular}
}
\end{table}

\subsubsection*{Example 2: Nonlinear models}
Consider the following four non-linear data generating models
\begin{description}
  \item[\quad \quad Model 2.a:] \ $Y = 5X_1 + 2\sin(\pi X_2/2) + 2 X_3 \bfm 1\{X_3 > 0\} + 2 \exp\{ 5X_4 \} + \varepsilon.$
  \item[\quad \quad Model 2.b:] \ $Y = 3X_1 + 3X_2^3 + 3X_3^{-1} + 5 \bfm 1\{X_4 > 0\} + \varepsilon.$
  \item[\quad \quad Model 2.c:] \ $Y = 1 - 5(X_2+X_3)^3\exp\{ -5(X_1 + X_4^2) \} + \varepsilon.$
  \item[\quad \quad Model 2.d:] \ $Y = 1 - 5(X_2+X_3)^{-3}\exp\{ 1 + 10 \sin(\pi X_1/2) + 5 X_4 \} + \varepsilon.$
\end{description}
Models 2.a and 2.b admit additive structure while Models 2.c and 2.d have more challenging nonlinear structures.
In addition, we generate $\bx \sim N(\bfm 0, \bfm\Sigma)$ and $\varepsilon \sim N(0, 1)$. Hence, for each model above, the active set contains the first 4 covariates in $\bx$.

The quantiles of the minimum model size that includes all 4 active features are presented in Table \ref{tab_additive}. Again, we observe that PC-Screen significantly outperforms the other three methods in all scenarios.
For Model 2.a, the $50\%$ quantile of the minimum model size of PC-Screen is exactly four while the other three methods need much larger model sizes to recover the active set. For Model 2.b, DC-SIS and bcDC-SIS preform comparably to PC-Screen at $5\%$ and $25\%$ quantiles but much worse for other higher quantiles.
For Models 2.c and 2.d, PC-Screen performs reasonably well while the other methods fail to effectively screen out the inactive features. The 95\% quantiles of SIS, DC-SIS, and bcDC-SIS are almost as large as $p$. This indicates, in the worst-case scenario, SIS, DC-SIS, and bcDC-SIS are hopeless to effectively reduce the dimensionality without missing any active feature.

\begin{table}[htp]
\centering
\caption{The quantiles of minimum model size for nonlinear models in Example 2 over 200 replications. The true model size is 4.}
\label{tab_additive}
\resizebox{1\textwidth}{!}{
\begin{tabular}{@{}rrrrrrrrrrrrrrr@{}}
\toprule
          && \multicolumn{5}{c}{\textbf{Model 2.a}}  & \phantom{ab} &\multicolumn{5}{c}{\textbf{Model 2.b}}  \\
          \cmidrule{3-7} \cmidrule{9-13}
          && 5\%       & 25\%   & 50\%   & 75\%   & 95\%   && 5\%       & 25\%   & 50\%   & 75\%   & 95\%   \\ \midrule
$p=5000$  &&           &        &        &        &        &&           &        &        &        &        \\
PC-Screen && 4.0       & 4.0    & 4.0    & 5.2    & 19.1   && 4.0       & 5.0    & 9.5    & 26.5   & 261.9  \\
DC-SIS    && 600.0     & 1880.2 & 2994.5 & 4052.2 & 4701.3 && 4.0       & 7.8    & 61.0   & 541.5  & 2310.9 \\
bcDC-SIS  && 488.4     & 1480.8 & 2863.0 & 3967.8 & 4855.9 && 4.0       & 6.0    & 21.5   & 88.0   & 893.8  \\
SIS       && 709.0     & 2065.8 & 3062.5 & 4160.0 & 4869.4 && 54.0      & 658.5  & 2692.5 & 4213.0 & 4829.1 \\
\midrule
$p=10000$ &&           &        &        &        &        &&           &        &        &        &        \\
PC-Screen && 4.0       & 4.0    & 4.0    & 5.0    & 31.0   && 4.0       & 5.8    & 13.0   & 48.8   & 393.9  \\
DC-SIS    && 664.0     & 3162.5 & 5655.5 & 7605.8 & 9490.1 && 4.0       & 13.8   & 86.0   & 843.2  & 5529.2 \\
bcDC-SIS  && 663.4     & 2605.8 & 5578.5 & 7745.2 & 9208.4 && 4.0       & 8.0    & 24.5   & 150.5  & 1403.9 \\
SIS       && 986.2     & 4048.2 & 6068.0 & 8298.5 & 9752.6 && 64.5      & 1193.8 & 4488.0 & 8139.5 & 9725.6 \\
\midrule
&& \multicolumn{5}{c}{\textbf{Model 2.c}}  & \phantom{ab} &\multicolumn{5}{c}{\textbf{Model 2.d}}  \\
           \cmidrule{3-7} \cmidrule{9-13}
          && 5\%      & 25\%   & 50\%   & 75\%   & 95\%   && 5\%      & 25\%   & 50\%   & 75\%   & 95\%   \\ \midrule
$p=5000$  &&          &        &        &        &        &&          &        &        &        &        \\
PC-Screen && 4.0      & 4.0    & 4.0    & 6.0    & 21.3   && 4.0      & 5.0    & 9.5    & 36.0   & 169.4  \\
DC-SIS    && 397.6    & 1536.8 & 2750.5 & 3930.0 & 4721.8 && 1639.1   & 3138.5 & 3851.5 & 4434.8 & 4902.4 \\
bcDC-SIS  && 196.5    & 1267.5 & 2774.0 & 4236.0 & 4986.2 && 605.5    & 1251.8 & 2073.5 & 2972.5 & 4209.1 \\
SIS       && 421.2    & 1615.8 & 2920.0 & 4057.0 & 4761.0 && 2065.0   & 3487.8 & 4083.5 & 4659.0 & 4950.3 \\
\midrule
$p=10000$ &&          &        &        &        &        &&          &        &        &        &        \\
PC-Screen && 4.0      & 4.0    & 4.0    & 7.2    & 25.0   && 4.0      & 5.0    & 13.0   & 62.5   & 297.4  \\
DC-SIS    && 668.8    & 3628.2 & 6243.5 & 7920.5 & 9531.6 && 3409.6   & 6380.8 & 7705.0 & 8756.5 & 9790.6 \\
bcDC-SIS  && 288.2    & 2006.5 & 4714.5 & 7370.2 & 9869.9 && 776.4    & 2567.8 & 4154.5 & 5517.8 & 8251.0 \\
SIS       && 760.9    & 3609.5 & 6155.5 & 8129.8 & 9577.1 && 3333.4   & 6387.8 & 8007.5 & 9252.8 & 9847.2 \\
\bottomrule
\end{tabular}
}
\end{table}

\subsubsection*{Example 3: Multivariate response models}

In this experiment, we investigate the performance of PC-Screen for multivariate response models. We omit SIS method in this experiment as it is not applicable to multivariate response problems. We generate $\by = (Y_1, Y_2)^{\T}$ from a bivariate normal distribution with conditional mean $\bmu_{Y|X} = (\mu_{1}(\bx), \mu_{2}(\bx))^{\T} $ and covariance matrix $\bSigma_{Y|X} = (\sigma_{ij})_{2\times 2}$, where $\sigma_{11} = \sigma_{22} =1$ and $\sigma_{12} = \sigma_{21} = \sigma(\bx)$. Following the setting in \cite{li2012feature}, we set $\btheta = (\bfm 2^{\T}_4, \bfm 0^{\T}_{p-4})^{\T}$ and generate $\mu_1(\bx)$, $\mu_2(\bx)$ and $\sigma(\bx)$ from the two models below.
\begin{description}
  \item[\quad \quad Model 3.a:] \ $\mu_{1}(\bx) = \exp\{2(X_1 + X_2)\}$, $\mu_{2}(\bx) = X_3 + X_4$ and $\sigma(\bx) = \sin(\bx^{\T}\btheta)$.
  \item[\quad \quad Model 3.b:] \ $\mu_{1}(\bx) = 2\sin(\pi X_1/2) + X_3 + \exp\{1+X_4\}$, $\mu_{2}(\bx) = X_1^{-2} + X_2$ and \\$\sigma(\bx) = (\exp\{\bx^{\T}\btheta\}-1) / (\exp\{\bx^{\T}\btheta\}+1)$.
\end{description}
The union of the active sets of $\mu_1(\bx)$, $\mu_2(\bx)$ and $\sigma_1(\bx)$  contains the first four covariates in $\bx$.
The simulation results are summarized in Table \ref{tab_multiy}. Again, the PC-Screen method performs strikingly well compared to the other two methods.

\begin{table}[htp]
\centering
\caption{The quantiles of minimum model size for multivariate response models in Example 3 over 200 replications. The true model size is 4.}
\label{tab_multiy}
\resizebox{1\textwidth}{!}{
\begin{tabular}{@{}rrrrrrrrrrrrrrr@{}}
\toprule
          && \multicolumn{5}{c}{\textbf{Model 3.a}}  & \phantom{ab} &\multicolumn{5}{c}{\textbf{Model 3.b}}  \\
          \cmidrule{3-7} \cmidrule{9-13}
          && 5\%       & 25\%  & 50\%   & 75\%   & 95\%   && 5\%       & 25\%   & 50\%   & 75\%   & 95\%   \\ \midrule
$p=5000$  &&           &       &        &        &        &&           &        &        &        &        \\
PC-Screen && 4.0       & 4.0   & 4.0    & 4.0    & 6.0    && 4.0       & 4.0    & 6.0    & 18.0   & 114.4  \\
DC-SIS    && 53.0      & 463.0 & 1211.0 & 2349.0 & 3774.2 && 641.1     & 2308.8 & 3307.0 & 4257.8 & 4838.0 \\
bcDC-SIS  && 24.0      & 215.5 & 758.5  & 1965.0 & 3999.8 && 225.2     & 1270.2 & 2494.5 & 3596.8 & 4709.0 \\
\midrule
$p=10000$ &&           &       &        &        &        &&           &        &        &        &        \\
PC-Screen && 4.0       & 4.0   & 4.0    & 4.0    & 9.0    && 4.0       & 4.0    & 8.0    & 30.2   & 302.9  \\
DC-SIS    && 136.8     & 978.5 & 2237.5 & 4506.0 & 8106.6 && 1804.7    & 4510.5 & 6445.5 & 8153.5 & 9707.6 \\
bcDC-SIS  && 83.0      & 546.0 & 1828.5 & 4264.5 & 7711.3 && 444.4     & 2217.0 & 4695.0 & 7122.5 & 9076.4 \\ \bottomrule
\end{tabular}
}
\end{table}


\subsection{FDR control performance}

In this subsection, we use simulated examples to numerically assess the FDR control as well as sure screening property of the proposed PC-Knockoff procedure. We refer to Algorithm \ref{alg:fs_knockoff} for implementation details.

\subsubsection*{Example 4: FDR control for linear and Poisson models}
Consider the following five regression models with ten active variables.
\begin{description}
  \item[\quad \quad Model 4.a:] 
  \ Same as Model 1.a except that $\bbeta=(\bfm 1_s, \bfm 0_{p-s})$ with $s=10$.
  \item[\quad \quad Model 4.b:] \ Same as Model 4.a except that $\varepsilon\sim t_2$, the $t$ distribution with degrees of freedom 2.
  \item[\quad \quad Model 4.c:] \ Same as Model 4.a except that $\bx = 0.9\bx_1 + 0.1\bx_2$, where $\bx_1$ and $\bx_2$ are independently drawn from $\bx_1 \sim N(\bfm 0, \bSigma)$ and  $\bx_2\sim t_2(\bfm 0, \bSigma)$, respectively.
  \item[\quad \quad Model 4.d:] 
  \ Same as Model 1.e except that $\bbeta=2\cdot(\bfm 1_s, \bfm 0_{p-s})$ with $s=10$.
  \item[\quad \quad Model 4.e:] 
  \ Same as Model 1.f except that $\bbeta=2\cdot(\bfm 1_s, \bfm 0_{p-s})$ with $s=10$.
\end{description}

In this example, we set $n=1000$, $p=5000$ and repeat 200 replications for each scenario. In each replication, we randomly divide the sample into two non-overlapping subsamples.
The sample size and target dimension in the screening step are set to be $n_1=250$ and $d=100$, and the sample size used to construct knockoff features is $n_2=750$.
In addition, the covariance matrix is set to be $\bSigma=(\sigma_{ij})$ with $\sigma_{ij}=0.5^{|i-j|}$. The performance of FDR control is examined under a sequence of specified levels: $\alpha = 0.1, 0.15, 0.20, 0.25$ and $0.30$.

We summarize the results in Table \ref{tab:sim_FDR} in which $\alpha$ is the pre-specified FDR level, $|\hat\calA|$ is the average number of selected variables, the column `$X_j$' represents the probability that the active variable $X_j$ is selected, the column `All' represents the sure screening probability (i.e., the probability that all active variables are selected) and the column `$\hat{\text{FDR}}$' is the empirical FDR (i.e., the average of empirical FDP).
According to Table \ref{tab:sim_FDR}, the proposed PC-Knockoff procedure controls the empirical FDR under the pre-specified level $\alpha$ for almost all scenarios. The only exception is in Model 4.c we have $\hat{\text{FDR}}=0.254$ at $\alpha=0.25$.
In Model 4.c, $\bx$ is drawn from a mixture of multivariate normal and multivariate $t_2$ distributions. As a result, the second-order knockoffs may not approximate the exact ones very well. For the other four models, the second-order knockoffs perform well as features are normally distributed. Besides FDR control, the PC-Knockoff procedure maintains the sure screening property reasonably well in all scenarios. For Models 4.a -- 4.d, the probabilities of selecting all active variables are ranging from $91.5\%$ to $99.5\%$ regardless of the pre-specified level $\alpha$. Even for the more challenging scenario Model 4.e, the probability of selecting all active variables simultaneously is greater than $93.5\%$ when $\alpha \geq 0.2$. Please notice that this is achieved under very small model sizes (i.e., small $|\hat\calA|$), thanks to the knockoff features.

\begin{table}[h]
\centering
\caption {FDR control for linear and Poisson models in Example 4. The true model size is 10.}
\label{tab:sim_FDR}
\resizebox{1\textwidth}{!}{
\begin{tabular}{@{}rcccccccccccccccccccc@{}}
\toprule
$\alpha$ & $|\hat\calA|$ & $X_1$ & $X_2$ & $X_3$ & $X_4$ & $X_5$ & $X_6$ & $X_7$ & $X_8$ & $X_9$ & $X_{10}$ & \text{All} & $\hat{\text{FDR}}$ \\
\midrule
\phantom{a} &\multicolumn{13}{c}{\textbf{Model 4.a}} \\
0.10 & 11.245 & 0.995 & 0.995 & 1.000 & 0.995 & 0.995 & 0.995 & 0.995 & 0.995 & 0.995 & 0.990 & 0.990 & 0.097 \\
0.15 & 11.935 & 0.995 & 1.000 & 1.000 & 1.000 & 1.000 & 1.000 & 1.000 & 1.000 & 1.000 & 0.995 & 0.990 & 0.130 \\
0.20 & 12.860 & 1.000 & 1.000 & 1.000 & 1.000 & 1.000 & 1.000 & 1.000 & 1.000 & 1.000 & 0.995 & 0.995 & 0.188 \\
0.25 & 14.095 & 1.000 & 1.000 & 1.000 & 1.000 & 1.000 & 1.000 & 1.000 & 1.000 & 1.000 & 0.995 & 0.995 & 0.244 \\
0.30 & 15.270 & 1.000 & 1.000 & 1.000 & 1.000 & 1.000 & 1.000 & 1.000 & 1.000 & 1.000 & 0.995 & 0.995 & 0.285 \\
\midrule
\phantom{a} &\multicolumn{13}{c}{\textbf{Model 4.b}} \\
0.10 & 10.545 & 0.945 & 0.945 & 0.960 & 0.950 & 0.955 & 0.950 & 0.955 & 0.955 & 0.955 & 0.945 & 0.915 & 0.079 \\
0.15 & 11.350 & 0.985 & 0.990 & 0.995 & 0.995 & 0.995 & 1.000 & 0.995 & 1.000 & 0.990 & 0.970 & 0.920 & 0.100 \\
0.20 & 12.460 & 0.990 & 0.995 & 1.000 & 0.995 & 0.995 & 1.000 & 0.995 & 1.000 & 1.000 & 0.985 & 0.955 & 0.164 \\
0.25 & 13.570 & 0.995 & 1.000 & 1.000 & 0.995 & 0.995 & 1.000 & 0.995 & 1.000 & 1.000 & 0.985 & 0.965 & 0.216 \\
0.30 & 14.705 & 1.000 & 1.000 & 1.000 & 0.995 & 0.995 & 1.000 & 1.000 & 1.000 & 1.000 & 0.990 & 0.980 & 0.260 \\
\midrule
\phantom{a} &\multicolumn{13}{c}{\textbf{Model 4.c}} \\
0.10 & 11.165 & 0.995 & 0.995 & 0.995 & 1.000 & 0.995 & 0.995 & 0.995 & 0.995 & 0.995 & 0.995 & 0.995 & 0.092 \\
0.15 & 11.660 & 1.000 & 0.995 & 1.000 & 1.000 & 1.000 & 1.000 & 1.000 & 1.000 & 1.000 & 0.995 & 0.995 & 0.116 \\
0.20 & 13.075 & 1.000 & 0.995 & 1.000 & 1.000 & 1.000 & 1.000 & 1.000 & 1.000 & 1.000 & 0.995 & 0.995 & 0.193 \\
0.25 & 14.420 & 1.000 & 0.995 & 1.000 & 1.000 & 1.000 & 1.000 & 1.000 & 1.000 & 1.000 & 0.995 & 0.995 & 0.254 \\
0.30 & 15.570 & 1.000 & 0.995 & 1.000 & 1.000 & 1.000 & 1.000 & 1.000 & 1.000 & 1.000 & 0.995 & 0.995 & 0.299 \\
\midrule
\phantom{a} &\multicolumn{13}{c}{\textbf{Model 4.d}} \\
0.10 & 10.630 & 0.940 & 0.945 & 0.940 & 0.945 & 0.955 & 0.940 & 0.950 & 0.940 & 0.950 & 0.940 & 0.925 & 0.092 \\
0.15 & 11.660 & 0.985 & 0.980 & 0.995 & 0.985 & 0.995 & 0.990 & 1.000 & 1.000 & 0.990 & 0.990 & 0.930 & 0.122 \\
0.20 & 12.870 & 0.990 & 0.990 & 1.000 & 1.000 & 1.000 & 0.995 & 1.000 & 1.000 & 0.990 & 0.990 & 0.965 & 0.193 \\
0.25 & 13.830 & 0.990 & 0.995 & 1.000 & 1.000 & 1.000 & 0.995 & 1.000 & 1.000 & 0.995 & 0.995 & 0.975 & 0.242 \\
0.30 & 15.020 & 0.995 & 0.995 & 1.000 & 1.000 & 1.000 & 0.995 & 1.000 & 1.000 & 0.995 & 0.995 & 0.980 & 0.288 \\
\midrule
\phantom{a} &\multicolumn{13}{c}{\textbf{Model 4.e}} \\
0.10 &  9.790 & 0.855 & 0.870 & 0.860 & 0.860 & 0.865 & 0.870 & 0.875 & 0.870 & 0.865 & 0.865 & 0.815 & 0.086 \\
0.15 & 11.590 & 0.955 & 0.985 & 0.975 & 0.965 & 0.980 & 0.995 & 0.980 & 0.970 & 0.965 & 0.970 & 0.825 & 0.123 \\
0.20 & 12.895 & 0.970 & 0.990 & 0.995 & 0.995 & 1.000 & 1.000 & 0.995 & 0.990 & 0.985 & 0.980 & 0.935 & 0.189 \\
0.25 & 14.025 & 0.980 & 0.995 & 0.995 & 0.995 & 1.000 & 1.000 & 0.995 & 0.990 & 0.990 & 0.990 & 0.935 & 0.244 \\
0.30 & 15.040 & 0.985 & 1.000 & 0.995 & 0.995 & 1.000 & 1.000 & 0.995 & 0.990 & 0.990 & 0.990 & 0.945 & 0.283 \\
\bottomrule
\end{tabular}
}
\end{table}

To illustrate how the two steps in PC-Knockoff work together, we also compare the active set $\hat\calA_1$ selected by the screening step without knockoff features (i.e. selected by PC-Screen) and the active set $\hat\calA(T_{\alpha})$ selected by PC-Knockoff. Table \ref{tab:delta_talpha} summarizes the average number of selected features $|\hat\calA|$, the sure screening probability `All' and empirical false discovery rate $\hat\FDR$ for Models 4.a -- 4.e. For the screening step, we conservatively select active features by including the top $d=100$ features. The results show that $\hat\calA_1$ enjoys the sure screening property but has a very high empirical FDR ($\hat\FDR = 0.90$). On the contrary, the active set $\hat\calA(T_\alpha)$ selected by the PC-Knockoff is able to control the FDR below the pre-specified level with only a little sacrifice of power.

\begin{table}[htp]
\centering
\caption{Comparison between $\hat\calA_1$ and $\hat\calA(T_\alpha)$. The number active features is $s=10$ and pre-specified FDR level for PC-Knockoff is $\alpha = 0.2.$}
\label{tab:delta_talpha}
\resizebox{0.7\textwidth}{!}{
\begin{tabular}{@{}cccccccccccccc@{}}
\toprule
& & \multicolumn{3}{c}{PC-Screen}  && \multicolumn{3}{c}{PC-Knockoff}  \\
\cmidrule{3-5} \cmidrule{7-9}
& & $|\hat\calA_1|$ & All  & $\hat{\FDR}$  && $|\hat\calA(T_{\alpha})|$ & All & $\hat{\FDR}$\\
\midrule
& \textbf{Model 4.a} & 100     & 0.995 & 0.90 && 12.9     & 0.995 & 0.188 \\
& \textbf{Model 4.b} & 100     & 0.995 & 0.90 && 12.5     & 0.955 & 0.164 \\
& \textbf{Model 4.c} & 100     & 0.995 & 0.90 && 13.0     & 0.995 & 0.193 \\
& \textbf{Model 4.d} & 100     & 0.995 & 0.90 && 12.9     & 0.965 & 0.193 \\
& \textbf{Model 4.e} & 100     & 0.995 & 0.90 && 12.9     & 0.935 & 0.189 \\
\bottomrule
\end{tabular}
}
\end{table}

\subsection{Supermarket data}

In this subsection, we apply the PC-Knockoff procedure to study a supermarket dataset \citep{wang2009forward,chen2018error}. The dataset consists of $n=464$ observations of daily records from a supermarket. The response $Y$ is the number of customers visited the supermarket on that day. The covariates are sale volumes of $p=6398$ products.  Due to data privacy, the detailed product codes are not released in the dataset. Instead, we name the covariates by their indices, i.e., $X_1, \ldots, X_{6398}$.  Both the response and predictors have been standardized to have zero mean and unit variance. The goal is to screen a parsimonious set of products whose sale volumes significantly contribute to the daily number of customers. Meanwhile, we want to control the FDR at level $\alpha=0.1$.

\begin{figure}[htp]
    \centering
    \includegraphics[width=\textwidth]{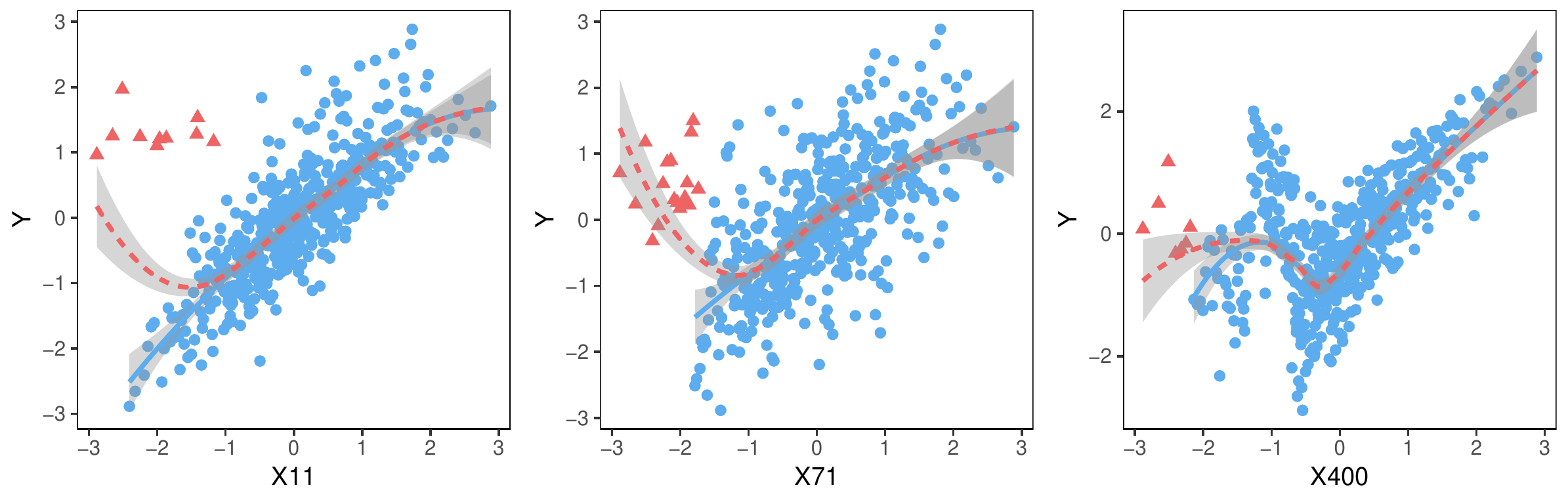}
    \caption{Scatter plots between the response and variables $X_{11}, X_{71}$ and $X_{400}$. The red triangles are the potential outliers and blue dots are the rest observations. The red dashed curves and blue solid curves are fitted local polynomial regression curves with and without the potential outliers, receptively. The gray shaded areas are corresponding confidence regions. }
    \label{fig:market_3}
\end{figure}

\begin{figure}[h]
    \centering
    \includegraphics[width=0.9\textwidth]{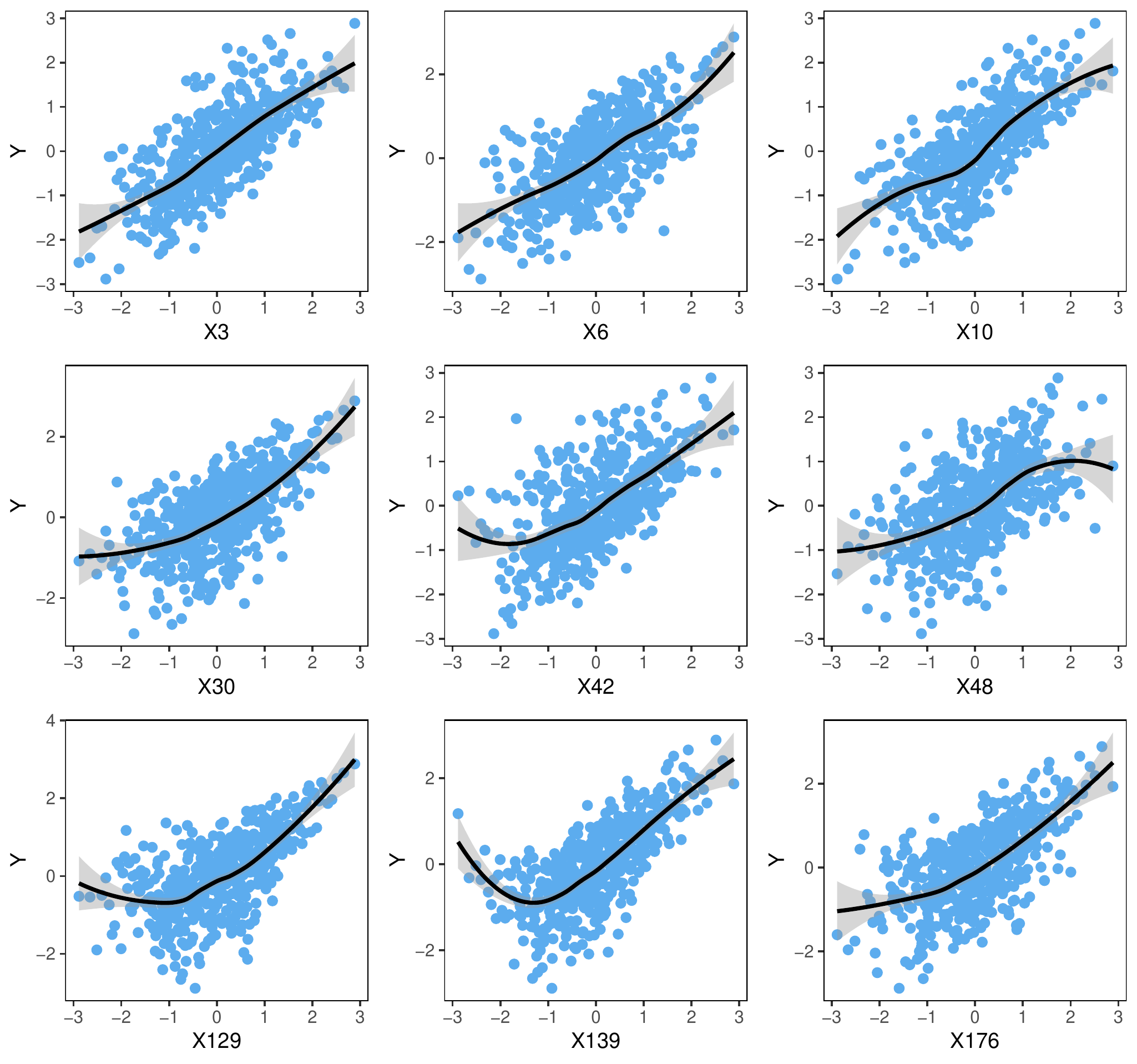}
    \caption{Scatter plots between the response and variables $X_{3}$, $X_{6}$, $X_{10}$, $X_{30}$, $X_{42}$, $X_{48}$, $X_{129}$, $X_{139}$ and $X_{176}$. The blue dots are observations. The black solid curves are fitted local polynomial regression curves. The gray shaded areas are corresponding confidence regions.}
    \label{fig:market_9}
\end{figure}

The implementation of PC-Knockoff follows Algorithm \ref{alg:fs_knockoff}. We set $n_1=200$ and $d=50$ in the screening step. That is, we randomly choose $200$ observations and use PC-Screen to pre-screen 50 features in the screening step. Then we construct second-order knockoffs for the pre-screened 50 features using the remaining 264 observations in the knockoff step. Under the pre-specified FDR level $\alpha=0.1$, the PC-Knockoff procedure selects a model of $12$ variables: $ X_3, X_6, X_{10}, X_{11}, X_{30}, X_{42}, X_{48}, X_{71}, X_{129}, X_{139}, X_{176}$, and  $X_{400}$. For each selected variable, we draw a scatter plot between this variable and the response. We observe obvious outliers in the scatter plots of  $X_{11}, X_{71}$ and $X_{400}$. We report this observation in Figure~\ref{fig:market_3}, where red triangles denote potential outliers and blue dots denote the rest data points. Besides, red dashed curves and blue solid curves are the fitted local polynomial regression curves with and without potential outliers, receptively. The gray shaded areas are the $95\%$ confidence regions. By comparing blue and red curves, we find the existence of outliers visually alters the fitted curves. This justifies the PC-Knockoff procedure is insensitive to the presence of outliers. The scatter plots between the response and other nine variables ($X_{3}$, $X_{6}$, $X_{10}$, $X_{30}$, $X_{42}$, $X_{48}$, $X_{129}$, $X_{139}$ and $X_{176}$) are presented in  Figure~\ref{fig:market_9}, which includes the local polynomial regression curves with the $95\%$ confidence regions. Figure~\ref{fig:market_9} indicates that PC-Knockoff can detect various functional relationships.

This supermarket dataset has also been analyzed by \cite{chen2018error}. They first apply DC-SIS to screen variables and then fit an additive model with selected variables. They further employ the Wald's $\chi^2$-test with the refitted cross-validation error variance estimate to determine if the selected features are significant at a pre-specified level. As a result, \cite{chen2018error} selected seven significant variables $X_{3}, X_{6}, X_{11}, X_{39}, X_{42}, X_{62}$ and $X_{139}$.
Next, we compare the in-sample fitting and out-of-sample prediction performance between the model selected by PC-Knockoff and the one in \cite{chen2018error} through a bootstrap experiment. In each replication, we randomly split the dataset into a training set of size 400 and a test set of size 64. We fit two additive models with the features selected by the two competitive methods, respectively. Then we calculate and record the training and testing $R^2$s for the two models. We repeat it for 200 replications. The sample mean and sample standard deviation of $R^2$s are reported in Table~\ref{tab:add_r2}, which show that the model selected by PC-Knockoff yields higher sample means of $R^2$ for both training set and test set.

\begin{table}[htp]
\centering
\caption{Sample mean and sample standard deviation of $R^2$ for training set and test set over 200 bootstrap replications for supermarket data.}
\label{tab:add_r2}
\begin{tabular}{lcccccccccccccc}
\toprule
 & \multicolumn{2}{c}{\textbf{Training set}} && \multicolumn{2}{c}{\textbf{Test set}} \\
\cmidrule{2-3} \cmidrule{5-6}
            & \textbf{Mean} & \textbf{SD} && \textbf{Mean} & \textbf{SD} \\
\midrule
PC-Knockoff & 0.8681 & 0.0042 && 0.8608 & 0.0281 \\
DC-SIS with $\chi^2$-test & 0.8534 & 0.0043 && 0.8488 & 0.0276 \\
\bottomrule
\end{tabular}
\end{table}

Due to the limited space, we present an additional real application to microarray data in Section S.2 of the supplementary file.

\section{Conclusion}\label{sec:con}
We study the feature screening problems under a general setup. The proposed PC-Screen method developed in Section \ref{sec:screen} is novel in the sense that it does not impose any regression model assumption and is robust against possibly heavy-tailed data. The response variable is also allowed to be multidimensional. Regarding the theoretical aspect, the asymptotic results developed in \cite{zhu2017projection} do not apply to the ``large $p$ small $n$ problem", and hence we develop non-asymptotic results for the empirical projection correlation.
The theoretical analysis shows that the PC-Screen method satisfies sure screening and rank consistency properties. Numerically, we show PC-Screen outperforms popular competitors over various data generative processes.
Moreover, this paper tackles the false discovery rate control problem in feature screening. In Section \ref{sec:FDR}, we propose a two-step procedure named PC-Knockoff to determine the threshold of active features. With the sample splitting idea, we first screen the ultra-high dimensional features to a moderate model size and then further select a model with FDR control using a statistic built upon knockoff features. The PC-Knockoff procedure controls FDR under a pre-specified level while maintaining sure screening property with high probability. In practice, the PC-Knockoff procedure works well in various scenarios.

\setcounter{equation}{0}
\setcounter{section}{0}
\setcounter{subsection}{0}

\section*{Appendix: Proof of results in Section \ref{sec:FDR}}\label{sec:app_main}

\baselineskip=20pt

\renewcommand{\theequation}{A.\arabic{equation}}
\renewcommand{\thetable}{A\arabic{table}}
\renewcommand{\thesubsection}{A.\arabic{subsection}}

In the appendix, we provide the proofs of Theorem~\ref{thm:pc_fdr} and Theorem~\ref{Thm_12} in Section \ref{sec:FDR}. We relegate the proofs of remaining results as well as some additional numerical examples to a supplementary file.

\subsection{Proof the Theorem~\ref{thm:pc_fdr}}

Denote $\hat\calA_1$, the set of features that are selected in the screening step. Throughout this proof, we restrict ourselves to the event ${\cal E} = \{ {\calA} \subset \hat\calA_1 \}$. According to Theorem~\ref{thm:ss}, the probability of $\cal E$ is at least $1 - O(s\exp\{-c_4n_1^{1-2 \kappa}\})$. To simplify the notations, we omit the condition on $\cal E$.

Denote by $\calB = \calA\cap\hat\calA_1$ and $\calB^c = \calA^c\cap\hat\calA_1$ the intersections between active and inactive sets with $\hat\calA_1$, respectively.
Notice that any feature with $|\hat W_j| = 0$ will not be included in the final selected set $\hat\calA(T_\alpha)$.
Hence, without loss of generality, we assume $\hat\calA_1 = \{1, 2, \dots, d\}$ and $|\hat W_1| \geq |\hat W_2| \geq \cdots \geq |\hat W_d| > 0$. For ease of presentation, we further define $\hat W_{d+1}=0$. Then we have
\begin{equation}
\begin{aligned}
       &\E\left[ \frac{\#\{ j:j\in \calB^c \cap \hat\calA(T_\alpha) \}}{\#\{ j:j\in\hat\calA(T_\alpha) \} \vee 1} \right] \\
      =& \E\left[ \frac{\#\{ j:j\in \calB^c \ \text{and} \ \hat W_j \geq T_\alpha \}}{\#\{ j:j\in\hat\calA(T_\alpha) \} \vee 1} \right] \\
      =& \E\left[ \frac{\#\{ j:j\in \calB^c \ \text{and} \ \hat W_j \geq T_\alpha \}}{1 + \#\{ j:j\in \calB^c \ \text{and} \ \hat W_j \leq -T_\alpha \}} \cdot \frac{1 + \#\{ j:j\in \calB^c \ \text{and} \ \hat W_j \leq -T_\alpha \}}{\#\{ j:j\in\hat\calA(T_\alpha) \} \vee 1} \right] \\
      \leq & \E\left[ \frac{\#\{ j:j\in \calB^c \ \text{and} \ \hat W_j \geq T_\alpha \}}{1 + \#\{ j:j\in \calB^c \ \text{and} \ \hat W_j \leq -T_\alpha \}} \cdot \frac{1 + \#\{ j: j \in \hat\calA_1 \ \text{and} \ \hat W_j \leq -T_\alpha \}}{\#\{ j:j\in\hat\calA(T_\alpha) \} \vee 1} \right] \\
      \leq & \E\left[ \frac{\#\{ j:j\in \calB^c \ \text{and} \ \hat W_j \geq T_\alpha \}}{1 + \#\{ j:j\in \calB^c \ \text{and} \ \hat W_j \leq -T_\alpha \}} \cdot \alpha \right].
\end{aligned}
\label{ineq:fdr}
\end{equation}
The first inequality holds since $\{ j:j\in \calB^c \ \text{and} \ \hat W_j \leq -T_\alpha \} \subset \{ j: \in \hat\calA_1 \ \text{and} \ \hat W_j \leq -T_\alpha \}$ and the second inequality is due to the definition of $T_\alpha$ in  \eqref{eq:ta}. In order to find $T_\alpha$, one can simply try different values of $t$ starting from the smallest value $t=|\hat W_{d+1}|=0$, then move to the second smallest value $t=|\hat W_d|$, then move to $t=|\hat W_{d-1}|$, and so on. The procedure stops as soon as it finds a value of $t$ satisfying \eqref{eq:ta}. In this process, $T_\alpha$ can be regarded as a stopping time. More rigorously, for $k = d+1, d, d-1, \dots, 1$, we define
\begin{equation*}
\begin{aligned}
      M(k) &= \frac{\#\{ j:j\in \calB^c \ \text{and} \ \hat W_j \geq |\hat W_k| \}}{1 + \#\{ j:j\in \calB^c \ \text{and} \ \hat W_j \leq -|\hat W_k| \}} \\
      &= \frac{\#\{ j:j\in \calB^c, j \leq k, \hat W_j > 0 \}}{1 + \#\{ j:j\in \calB^c, j \leq k, \hat W_j \leq 0 \}} \\
      &=: \frac{V_+(k)}{1 + V_-(k)},
\end{aligned}
\end{equation*}
where $V_+(k) = \#\{ j:j\in \calB^c, j \leq k, \hat W_j > 0 \}$ and $V_-(k) = \#\{ j:j\in \calB^c, j \leq k, \hat W_j \leq 0 \}$.
Let $\calF_k$ be the $\sigma$-algebra generated by $\{ V_\pm(d+1),V_\pm(d), \dots, V_\pm(k), Z_{d+1}, Z_d\dots, Z_k \}$ where $Z_{d+1} = 0$ and
\begin{equation*}
      Z_j =
      \begin{cases}
            1 \quad \text{if} \ j \in \calB \\
            0 \quad \text{if} \ j \in \calB^c. \\
      \end{cases}
\end{equation*}
As a result, given $\calF_k$, we know whether $k$ is in the active set $\calB$ or not.

Next we show that the process $M(d+1), M(d), \dots, M(1)$ is a super-martingale running backward with respect to the filtration $\calF_{d+1} \subset \calF_{d} \subset \dots \subset \calF_1$. On one hand, if $k \in \calB$, we have $V_+(k) = V_+(k-1)$, $V_-(k)=V_-(k-1)$ and thus $M(k) = M(k-1)$. On the other hand, if $k \in \calB^c$, we have
\[
      M(k-1) = \frac{V_+(k)-I_k}{1 + V_-(k)-(1-I_k)} = \frac{V_+(k)-I_k}{(V_-(k) + I_k)\vee 1},
\]
where $I_k = I\{W_k > 0\}$ and $I\{\cdot\}$ is the indicator function. Let $d_0 = \#\{ j:j\in \calB^c \}$ with the inactive set $\calB^c = \{j_1, j_2, \dots, j_{d_0}\}$. From Lemma~1 part (ii), we know that $I_{j_1}, I_{j_2} \dots, I_{j_{d_0}}$ are i.i.d. \text{Bernoulli}(0.5) random variables. Thus conditioning on $\calF_k$, (hence $V_+(k)$, $V_-(k)$ are known), we have
\[
      \text{Pr}(I_k = 1|\calF_k) =  \text{Pr}(I_k = 1|V_+(k),V_-(k)) = \frac{V_+(k)}{V_+(k) + V_-(k)}.
\]
Thus in the case $k \in \calB^c$,
\begin{equation*}
\begin{aligned}
      \E[M(k-1)|\calF_k] &=  \frac{V_+(k)}{V_+(k) + V_-(k)}\cdot \frac{V_+(k)-1}{V_-(k)+1} +  \frac{V_-(k)}{V_+(k) + V_-(k)}\cdot \frac{V_+(k)}{V_-(k)\vee 1} \\
      &=
      \begin{cases}
            \frac{V_+(k)}{V_-(k) + 1}, \ &\text{if} \ V_-(k)>0, \\
            V^+(k) - 1, \ &\text{if} \ V_-(k)=0. \\
      \end{cases} \\
      &=
      \begin{cases}
            M(k), \ &\text{if} \ V_-(k)>0, \\
            M(k) - 1, \ &\text{if} \ V_-(k)=0. \\
      \end{cases}
\end{aligned}
\end{equation*}
Therefore, $\E[M(k-1)|\calF_k] \leq M(k)$, implying that $M(k),k=d+1, \dots, 1$ is a super-martingale with respect to $\{\calF_k\}$. By definition, $T_\alpha$ is a stopping time with respect to the backward filtration $\{\calF_k\}$. According to the optional stopping theorem for super-martingale, we know
\[
      \E[M(k_{T_\alpha})] \leq \E[M(k_{d+1})] =
      \E\left[ \frac{\#\{ j:j\in \calB^c, W_j > 0 \}}{1 + \#\{ j:j\in \calB^c, W_j \leq 0 \}} \right] =
      \E\left[\frac{X}{1+d_0-X}\right],
\]
where $X=\#\{ j:j\in \calB^c, W_j > 0 \}$. Since $X \sim \text{Binomial}(d_0, 1/2)$, we have
\begin{equation*}
\begin{aligned}
      \E\left[\frac{X}{1+d_0-X}\right]
      =& \sum_{k=1}^{d_0} \binom{d_0}{k} \left(\frac{1}{2}\right)^{k} \left(\frac{1}{2}\right)^{d_0-k} \cdot \frac{k}{1+d_0-k} \\
      =& \sum_{k=1}^{d_0} \binom{d_0}{k-1} \left(\frac{1}{2}\right)^{k} \left(\frac{1}{2}\right)^{d_0-k} \\
      =& \sum_{k=0}^{d_0-1} \binom{d_0}{k} \left(\frac{1}{2}\right)^{k+1} \left(\frac{1}{2}\right)^{d_0-k-1} \\
      \leq& 1.
\end{aligned}
\end{equation*}
Therefore $\E[M(k_{T_\alpha})] \leq 1$.
From (\ref{ineq:fdr}), we have
\[
      \E\left[ \frac{\#\{ j:j\in \calB^c \cap \hat\calA(T_\alpha) \}}{\#\{ j:j\in\hat\calA(T_\alpha) \} \vee 1} \right] \leq \alpha \E[M(k_{T_{\alpha}})] \leq \alpha.
\]
Since $\hat\calA(T_\alpha) \subset \hat\calA_1$, we have $\#\{ j:j\in \calA^c \cap \hat\calA(T_\alpha) \}
= \#\{ j:j\in \calB^c \cap \hat\calA(T_\alpha) \}$. Then, we conclude
\[
      \E\left[ \frac{\#\{ j:j\in \calA^c \cap \hat\calA(T_\alpha) \}}{\#\{ j:j\in\hat\calA(T_\alpha) \} \vee 1} \right] \leq \alpha.
\]

\subsection{Proof of Theorem~\ref{Thm_12}}
\label{sec:proof_s_a}
Now we restrict ourselves to the subset $(\bX^{(2)}, \bY^{(2)})$.
For $j\in \hat\calA_1$, define $\hat\omega_j = \hat\pc(\bX^{(2)}_j, \bY^{(2)})^2$, $ \tilde \omega_j = \hat\pc(\tilde \bX^{(2)}_j, \bY^{(2)})^2$, $\omega_j = \pc(X_j, \by)^2$, and $\omega_j' = \pc(\tilde X_j, \by)^2$.
Theorem~\ref{thm:pc} implies that
\[
\begin{aligned}
       \text{Pr}(|\hat \omega_j - \omega_j| &> c_3n_2^{-\kappa}) \leq 5c_1\exp\{-c_4n_2^{1-2 \kappa}\} \ \text{and} \\
       \text{Pr}(|\tilde \omega_j - \omega'_j| &> c_3n_2^{-\kappa}) \leq 5c_1\exp\{-c_4n_2^{1-2 \kappa}\}.
\end{aligned}
\]
Thus we have
\begin{equation}
\begin{aligned}
      &\text{Pr}(|\hat W_j - W_j| > 2c_3n_2^{-\kappa}) \\
      =& \text{Pr}(|\hat \omega_j - \tilde \omega_j - (\omega_j - \omega'_j)|> 2c_3n_2^{-\kappa} ) \\
      \leq&  \text{Pr}(|\hat \omega_j - \omega_j| > c_3n_2^{-\kappa}) + \text{Pr}(|\tilde \omega_j - \omega'_j| > c_3n_2^{-\kappa}) \\
      =&  O(\exp\{ -c_4 n_2^{1-2\kappa} \}).
\end{aligned}
\label{eqn:error_W_hat2}
\end{equation}
On the other hand, part (i) in Lemma~\ref{lem:null_wj} implies that $\omega_j = \omega_j'$ for all $j \in \calA^c$. We have
\begin{equation}
\begin{aligned}
&\text{Pr} \left( \underset{j\in \calB^c}{\max} |\hat W_j| \leq 2c_3 n_2^{-\kappa} \right )\\
            =& \text{Pr} \left( \underset{j\in \calB^c}{\max} |\hat \omega_j - \omega_j + \omega'_j - \tilde \omega_j| \leq 2c_3 n_2^{-\kappa} \right ) \\
            \geq& \text{Pr} \left( \underset{j\in \calB^c}{\max} |\hat \omega_j - \omega_j| \leq c_3 n_2^{-\kappa} \ \text{and} \ \underset{j\in \calB^c}{\max} |\tilde\omega_j - \omega_j| \leq c_3 n_2^{-\kappa} \right) \\
            \geq & 1 - \sum_{j \in \calB^c} \text{Pr} \left(|\hat \omega_j - \omega_j| \geq c_3 n_2^{-\kappa} \right) - \sum_{j \in \calB^c} \text{Pr} \left(|\tilde \omega_j - \omega_j| \geq c_3 n_2^{-\kappa} \right) \\
            \geq & 1 - O(n_2 \exp\{ -c_4 n_2^{1-2\kappa} \}),
\end{aligned}
\label{eqn:error_W_hat1}
\end{equation}
where the last equality is implied by Theorem~\ref{error_bound} and the assumption $n_2>d$.
Since $\min_{k\in\calB} W_k \geq 4c_3n_2^{-\kappa}$ and $d < n_2$, (\ref{eqn:error_W_hat2}) implies that $\min_{k\in\calB} \hat W_k \geq 2c_4n_2^{-\kappa}$ with probability at least $1 -  O(n_2\exp\{ -c_4 n_2^{1-2\kappa} \})$. Together with (\ref{eqn:error_W_hat1}), we know that $\min_{j \in \calB}\hat W_j > \max_{j \in \calB^c} |\hat W_j|$ with probability $1 -  O(n_2\exp\{ -c_4 n_2^{1-2\kappa} \})$. In other words, with probability at least $1 -  O(n_2\exp\{ -c_4 n_2^{1-2\kappa} \})$, the active features will be ranked ahead of the inactive features. Recall that in order to find the cutoff value $T_\alpha$, we start from the smallest value $t=|\hat W_{d+1}|=0$, then move to the second smallest value $t=|\hat W_d|$, then move to $t=|\hat W_{d-1}|$, and so on. The procedure stops once it finds a value of $t$ satisfying  \eqref{eq:ta}. Restrict on the event $\calE' = \{ \min_{j \in \calB}\hat W_j > \max_{j \in \calB^c} |\hat W_j| \}$, which holds with probability at least $1 -  O(n_2\exp\{ -c_4 n_2^{1-2\kappa} \})$ and let $t_{\min} = \min_{j \in \calB} |\hat W_j| = \min_{j \in \calB} \hat W_j$. If $1/s \leq \alpha$, then
      \[
            \frac{1+\#\{j: \hat{W}_j \leq -t_{\min} \}}{\#\{j: \hat{W}_j \geq t_{\min} \}} = \frac{1 + 0}{s} \leq \alpha.
      \]
      From the inequality above, we known this process must stop no later than $t$ reaches $t_{\min}$ and hence $T_\alpha \leq t_{\min}$. As a result,
      \[
            \hat\calA(T_\alpha) = \{j: \hat W_j \geq T_\alpha\} \supseteq \{j: \hat W_j \geq t_{\min}\} = \calA.
      \]
If $1/s > \alpha$, in order to satisfy  \eqref{eq:ta}, we must have $T_\alpha < t_{\min}$ or $T_\alpha > \max_j|\hat W_j|$.  $T_\alpha < t_{\min}$ indicates all active features are selected and $T_\alpha > \max_j|\hat W_j|$ leads to an empty $\hat\calA(T_\alpha)$, as stated in the second part. Notice that $|\hat W_{s+k}|$ satisfies the rule (\ref{eq:ta}) is equivalent to that at most $\lfloor (k-1)/(s+1)  \rfloor$ of the signs of $\{ \hat W_{s+1}, \dots, \hat W_{s+k}\}$ are negative. Let $a_k$ be the probability that $|\hat W_{s+k}|$ satisfies (\ref{eq:ta}) and $b_k$ be the probability that the process stops at $|\hat W_{s+k}|$ for $k = 1, \dots, d-s$. Let $a_0=0$, we have
$$
    a_k = \sum_{i=0}^{\lfloor (k-1)/(s+1)  \rfloor} {k \choose i}\left(\frac{1}{2}\right)^k, \\
    b_k = a_k(1-a_{k-1}- \cdots -a_0).
$$
It is easy to verify that for $s > 2$,
$$
\begin{aligned}
    &\prob(\calA \subset \hat\calA(T_\alpha)|\calE') = \sum_{k=1}^{d-s} b_k
    = \sum_{k=1}^{d-s} a_k(1-a_{k-1}-\cdots -a_0)\\
    &\leq \sum_{k=1}^{\infty} a_k(1-a_{k-1}-\cdots-a_0)= C(s) < 1,
\end{aligned}
$$
where $0 < C(s) < 1$ is some constant only depending on $s$. This completes the proof.

{\bf Acknowledgment}: We would like to thank the AE and reviewers for their 
constructive comments, which lead to significant improvement of this work.
Liu's research was supported by National Natural Science Foundation of China
(No. 11771361, 11671334, 11871409), Basic Scientific Center Project 71988101 of National Science Foundation of China  and JAS14007. Li's research was supported by
NSF grants DMS 1820702 and 1953196. All authors equally contributed to this work, and the authors are listed in seniority. Both Yuan Ke and Jingyuan Liu are designated to be the corresponding authors of this paper.

\baselineskip=15pt

\renewcommand{\baselinestretch}{1.0}

\bibliographystyle{agsm}
\bibliography{PC_Screen}

\end{document}